\documentclass[
	aps, twocolumn, amsmath, amssymb,
	prl, superscriptaddress, floatfix]{revtex4}

\usepackage{graphicx}		
\usepackage{dcolumn}		
\usepackage{bm}				
\usepackage{amssymb}
\usepackage[hypertexnames=false]{hyperref}
\usepackage{multirow}
\usepackage{color}
\usepackage[normalem]{ulem}

\newcommand{\tx}[1]{{\text{#1}}}
\newcommand{\bra}[1]{{\langle #1 \rangle}}

\begin{document}
 

\title{Dynamics of Motility-Induced clusters: coarsening beyond Ostwald ripening }

\date{\today}
\author{Claudio B. Caporusso}
\affiliation{Dipartimento  di  Fisica,  Universit\`a  degli  Studi  di  Bari  and  INFN,
    Sezione  di  Bari,  via  Amendola  173,  Bari,  I-70126,  Italy}

\author{Leticia F. Cugliandolo}
\affiliation{Sorbonne Universit\'e, Laboratoire de Physique Th\'eorique et Hautes Energies, CNRS UMR 7589,
    4 Place Jussieu, 75252 Paris Cedex 05, France}
\affiliation{Institut Universitaire de France, 1 rue Descartes, 75005 Paris France}

\author{Pasquale~Digregorio}
\affiliation{Dipartimento  di  Fisica,  Universit\`a  degli  Studi  di  Bari  and  INFN,
    Sezione  di  Bari,  via  Amendola  173,  Bari,  I-70126,  Italy}
\affiliation{CECAM  Centre  Europ\'een  de  Calcul  Atomique  et  Mol\'eculaire,
    Ecole  Polytechnique  F\'ed\'erale  de  Lausanne,  Batochimie,  Avenue  Forel  2,  1015  Lausanne,  Switzerland}

\author{Giuseppe Gonnella}
\affiliation{Dipartimento  di  Fisica,  Universit\`a  degli  Studi  di  Bari  and  INFN,
    Sezione  di  Bari,  via  Amendola  173,  Bari,  I-70126,  Italy}

\author{Demian Levis}
\affiliation{Departement de Fisica de la Materia Condensada, Facultat de Fisica, Universitat de Barcelona, Mart\'{\i}  i  Franqu\`es  1,  E08028  Barcelona,  Spain}
\affiliation{UBICS  University  of  Barcelona  Institute  of  Complex  Systems,  Mart\'{\i}  i  Franqu\`es  1,  E08028  Barcelona,  Spain}

\author{Antonio Suma}
\affiliation{Dipartimento  di  Fisica,  Universit\`a  degli  Studi  di  Bari  and  INFN,
    Sezione  di  Bari,  via  Amendola  173,  Bari,  I-70126,  Italy}

\email[email: ]{name@}
\begin{abstract}
We study the dynamics of clusters of Active Brownian Disks generated by Motility-Induced Phase Separation, 
by applying an algorithm that we devised to track cluster trajectories. 
    We identify an aggregation mechanism that goes beyond Ostwald ripening but also
    yields $z=3$.
    Active clusters of mass $M$ self-propel with enhanced diffusivity 
    $D\sim$ Pe$^2/\sqrt{M}$. Their fast motion drives aggregation into large fractal structures,
    which are  patchworks of diverse hexatic orders, and coexist with regular, orientationally uniform, smaller ones.
    To bring out the impact of activity, we perform a comparative study of a passive system
    that evidences major differences with the active case.
\end{abstract}

\maketitle
\vspace{0.25cm}


A conservative system of attractive particles  cooled across a condensation transition
undergoes particle aggregation: dense clusters or `droplets' grow following Ostwald ripening
at sufficiently long time scales. 
The theory of this process, due to Lifshitz, Slyozov and Wagner 
(LSW)~\cite{lifshitz1961kinetics,wagner1961theory, voorhees1985theory}, predicts an asymptotic growth of the clusters'
typical length, $R(t)\sim t^{1/3}$.

Understanding the dynamics across phase transitions
has been a central problem of condensed matter theory, partly due to the fact that it may allow one to define dynamic universality classes \cite{HohenbergRev,OnukiBook,BrayRev}.
In the context of active matter \cite{BechingerRev, WinklerRev, TailleurRev}, a key open question is how such universality classes, 
if any, are affected by non-equilibrium forces, e.g. self-propulsion.

Systems of repulsive self-propelled particles
experience Motility-Induced Phase Separation (MIPS). This phenomenon bears analogies with the phase separation of attractive passive particles even though, being purely triggered by activity, arises under non-equilibrium conditions~\cite{TailleurCates2008, CatesRev}.  In the quest to identify universal properties of MIPS, much work has been devoted to the analysis of simple particle models, such as  Active Brownian Particles (ABP) \cite{WinklerRev, romanczuk2012active, Fily2012, Redner2013f, Stenhammar2014, Wysocki, Caporusso20,PRLino, Caprini20, Gnan21,Valeriani21,PicaCiamarra22}, as well as field theories, in the spirit of the Hohenberg-Halperin classification \cite{HohenbergRev, cates2019active, Stenhammar13, wittkowski2014, Speck14, Speck15, Tjhung18, nardini2017entropy, caballerocates, paoluzzi2022scaling}.
The continuum theories, albeit  include terms that break time reversal symmetry,
predict the same growth law as in  phase separation under
conservative forces~\cite{cates2019active,Stenhammar13,wittkowski2014}, since those terms
turn out to be irrelevant in the renormalization group sense \cite{caballerocates,paoluzzi2022scaling}.
Accordingly, the kinetics of MIPS was also ascribed to  Ostwald ripening.
ABPs, when relaxed  at sufficiently large activities from a homogeneous state,  progressively form clusters and, after a transient,
enter a scaling regime in which
global structural measurements
scale with a growing length compatible with
$R(t) \sim t^{1/3}$~\cite{Redner2013f, Stenhammar2014, Caporusso20}, in line with universality. 

Dynamic clustering is a common feature of active colloids \cite{Cecile2012,Buttinoni13,Ginot2018,vanderLinden19}. Experimentally, clustering might result from more involved mechanisms than the mere competition between self-propulsion and excluded volume.
In fact, aggregation-fragmentation \cite{Ginot2018} and interrupted coarsening \cite{vanderLinden19} have been exhibited,
and both go beyond the LSW theory.

In this Letter we elucidate the role played by cluster motion
in the MIPS of ABPs.
With the cluster tracking method that we conceived, we show that the mechanisms in action
are not just Ostwald ripening. Indeed, cluster diffusion is a key player in the growth process, as it leads to cluster 
aggregation, another mechanism at work, even at long times.
Clusters break up and recombine and, pushed by the particles at their surface,
are much more mobile than their passive counterparts. Their enhanced motility is
evidenced by a diffusion coefficient
with anomalous mass dependence, $D\sim {\rm Pe}^2/\sqrt{M}$. 
Very rich cluster  structures are
thus built dynamically.
With a parallel study of passive attractive particles, we exhibit qualitative and quantitative differences with equilibrium phase separation.

\begin{figure}[b!]
    \centering
    \hspace{1.5cm}
    \includegraphics[width=8.3cm]{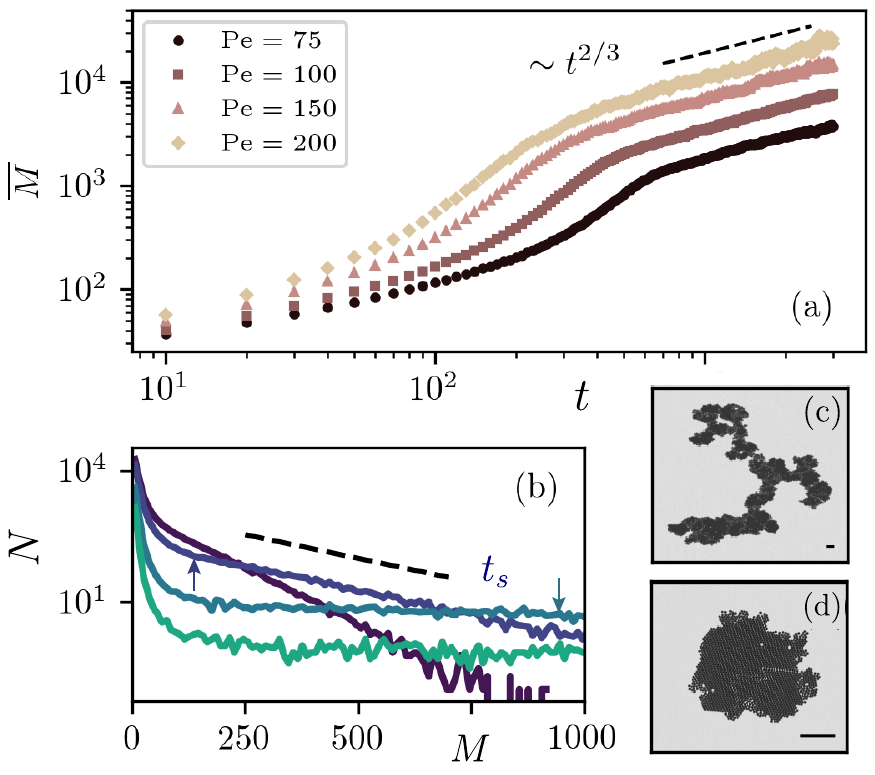}
    \caption{{\bf Cluster masses.}
        (a) Time dependence of their average, $\overline M$, for
        various Pe.
        In the scaling regime, reached after $t_s \sim $ Pe$^{-1}$, $\overline M(t) \sim t^{2/z}$ with an estimated dynamic exponent $z=3$.
        (b) Number of clusters of mass $M$ at $t=100,  300~(\sim t_s), 500, 1000$ (from dark violet to light green) for Pe = 100.
        The vertical arrows show the averaged mass $\overline M$ at $t = 300, \, 500$.
        The dashed line helps visualizing the exponential decay at $t_s$.
        (c) A large ($M\approx 2 \times 10^4$) and (d) a small ($M=1751$) typical cluster plotted with different scales (the two segments of length
        $10 \, \sigma_d$).
    }
    \label{fig:uno}
\end{figure}

We consider  $N=1024^2$
Active Brownian Particles (ABP) at evolving positions ${\bold{r}}_i$
in an $L_x\times L_y$ rectangular box with periodic boundary conditions:
\begin{equation}
    \begin{split}
        \label{eq:langevin}
        m\ddot{\bold{r}}_i+ \gamma\dot{\bold{r}}_i
        &
        =
        F_{\rm act} \bold{n}_i- \! \sum\limits_{j(\neq i)} \! {\boldsymbol{\nabla}}_iU(r_{ij})
        + \sqrt{2 \gamma k_B T}\bm{\xi}_i \; ,
        \\
        \dot{\theta}_i
        & =
        \sqrt{2 D_{\theta}}\eta_i
        \;  .
    \end{split}
\end{equation}
$F_{\rm act}$ is the modulus of the self-propulsion force acting along the direction
$\bold{n}_i=( \cos{\theta_i},\sin{\theta_i})$, $r_{ij}=|{\bold{r}}_i-{\bold{r}}_j|$ the inter-particle distance,
and $U(r)$ a repulsive Mie potential,
$
    U(r)=4\varepsilon [({\sigma}/{r})^{64}-({\sigma}/{r})^{32}]+\varepsilon
$
if $r< \sigma_d=2^{1/32}\sigma$ and $0$ otherwise.
The  components of $\bm{\xi}$ and $\eta$  are zero-mean  and unit variance independent white Gaussian noises.
The control parameters are the packing fraction
$\phi =\pi{\sigma^2_d}N/(4L_xL_y)$, with $L_x/L_y=2/\sqrt{3}$,
and the P\'eclet number Pe = $F_{\rm act} {\sigma_d}/(k_BT)$.
The persistence time $\tau_p=1/D_{\theta}=\sigma^2_d \gamma/(3 k_BT)$ relates to
the persistence length $l_p = F_{\rm act} \tau_p/\gamma$ = Pe$\,\sigma_d$/3
which grows proportionally to Pe.
Lengths and energies are measured in units of $ \sigma_d$ and $\varepsilon$, respectively.
We fix $\gamma=10$, $m=1$ and $k_BT=0.05$, consistently with
an over-damped limit and we show results for $\phi=0.5$.
We performed  extensive simulations,
starting from random initial conditions, and we devised and applied a
method, which tracks the clusters' trajectories individually.
We repeated our measurements in a system of passive
attractive particles interacting {\it via} the same, though not truncated,
Mie potential. Further details on the simulations can be found in the SM~\cite{SM}.


We start our analysis by monitoring the evolution of the averaged cluster mass (or size) and mass distribution. Figure~\ref{fig:uno}(a)
shows the growth of the averaged cluster mass (number of particles in the cluster)
$\overline M = \int dM \, P(M) \, M$, with
$P(M)$ the instantaneous mass distribution displayed (un-normalized) in Fig.~\ref{fig:uno}(b)
for Pe = 100. The scaling regime is
reached at the shoulder,  at a time $t_s\sim\text{Pe}^{-1}$  when $\overline M_s \sim$ Pe, Fig.~\ref{fig:scaling}(a).
In the scaling regime $\overline M(t)\sim t^{2/3}$,
consistently with the expectation  that the typical  length grows as $R(t)\sim t^{1/3}$. The curves
are successfully scaled as $\overline M(t)/\overline M_s = f(t/t_s)$,
Fig.~\ref{fig:scaling}(b).
The mass histogram $N(M)$ has a rapid exponential  decay at short times $t\leq  t_s$  over the full range of $M$.
Similar distributions of rather small clusters were measured
experimentally in  dilute samples~\cite{Ginot2018}.
In the scaling regime, $N(M)$ changes functional form
with,  in the selected range, a long plateau for $M \gtrsim 100$, and a further exponential decay, 
see Fig.~\ref{fig:CSDs} and its discussion~\cite{SM}.
The system is thus populated by many large clusters of different size and different geometry
(to be later quantified) that we illustrate in Fig.~\ref{fig:uno}(c)-(d).

\begin{figure}[t!]
    \centering
    \includegraphics[width=8.5cm]{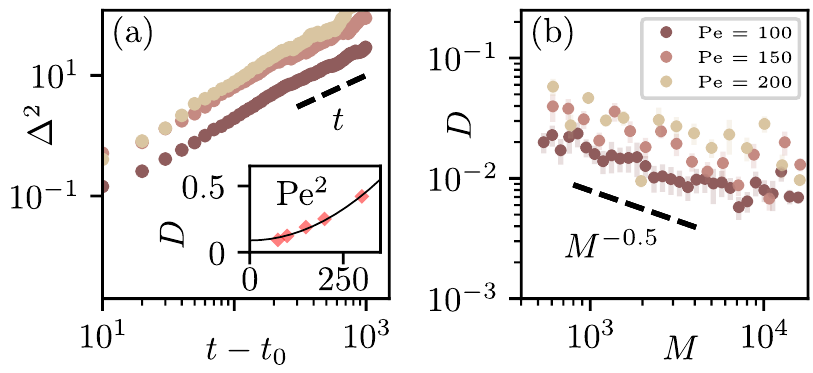}
    \vspace{-0.25cm}
    \caption{{\bf Cluster dynamics in the bulk.}
    (a) Average over all clusters' c.o.m. mean-square displacements, at three Pe,
    with  $t_0>t_s$ ($t_0 = 900$, $700$ and $500$ for $\rm{Pe}=100,150,200$, respectively).
    Inset: Pe dependence of the diffusion constant  together with the
    fit $T/\gamma \; (1+c {\rm Pe}^2)$ with $c = 8.1 \times 10^{-5}$ and
    $T/\gamma$ given by the external parameters.
    (b) Mass dependence of the single clusters' diffusion constant. $\overline M(t_0)$ are between $10^3$
    and $5 \times 10^3$ for all Pe.}
    \label{fig:dos}
\end{figure}

To better grasp the growth mechanism, and identify similarities and differences with the one acting in passive systems,
we turn our attention to the cluster's motion.

We compute the clusters' center of mass (c.o.m.) mean-\-square displacements, and we average over all the 
${\mathcal N}(t)$ clusters followed until time $t$: $\Delta^2 \equiv
        {\mathcal N}^{-1} \sum_{\alpha=1}^{{\mathcal N}} \Delta_\alpha^2$, with $\Delta^2_\alpha \equiv
    ({\mathbf r}^{\rm cm}_\alpha(t)-{\mathbf r}^{\rm cm}_\alpha(t_0))^2$,  and ${\mathbf r}^{\rm cm}_\alpha(t)$
the instantaneous position of the c.o.m. of the $\alpha$th cluster in the scaling regime,  $t_0>t_s$. The results are plotted in~Fig.~\ref{fig:dos}(a).
The earlier time ballistic and thermal diffusive scales are not captured with this measurement but one
can access them {\it via} the instantaneous cluster velocity (see  Fig.~\ref{fig:cluster_v2}).
The initial slope of $\Delta^2$ indicates super diffusion though not as pronounced as
for a single ABP. Later, {\it enhanced diffusion} sets in,
$\Delta^2 \simeq 2dD (t-t_0)$. The diffusion coefficient (insert) satisfies
$\gamma D = T_{\rm eff} =  T (1+ c {\rm Pe}^2)$, with the same functional form as for a single ABP,
though taking a considerably smaller value ($c = 8.1 \times 10^{-5}$ compared to 1/6  \cite{Fily2012}).
This result is compatible with the measurement of the effective temperature of the full dense component~\cite{Petrelli20},
with a comparable value $T_{\rm eff} \simeq 0.08$ for these parameters.
The individual $\Delta_\alpha^2$, though noisy and persistent, demonstrate that
the trajectories are Brownian
when examined at sufficiently long $t-t_0$.
The diffusion constants decrease with the clusters' mass and, although quite disperse, follow
$D(M) \sim M^{-\alpha}$ with $\alpha= 0.5$ (b). Data for $\phi=0.35$ are fitted with a similar $\alpha$.
We next argue about the origin of this enhanced diffusion and how it  affects the clusters' growth.

\begin{figure}[b!]
    \centering
    \includegraphics[width=8.5cm]{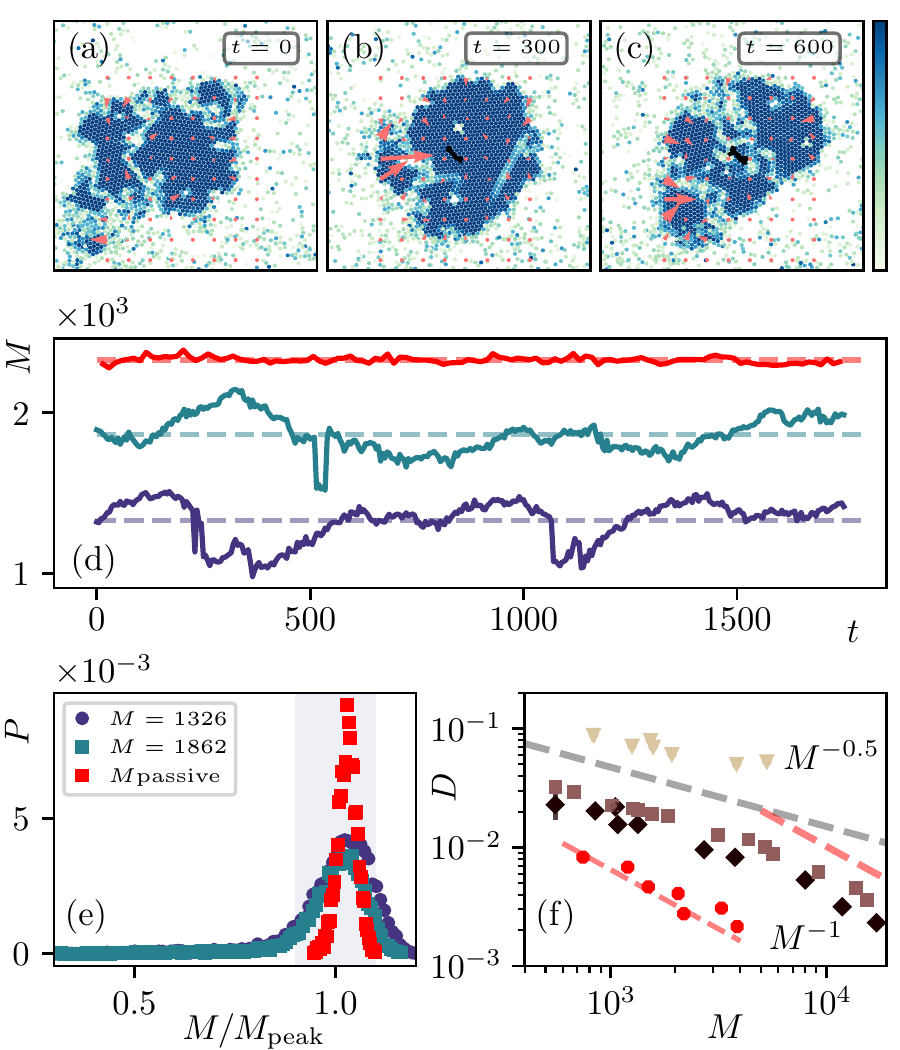}
    \caption{{\bf Dynamics of isolated  clusters.}
        (a)-(c) Subsequent snapshots of the same active cluster.
        The color code represents the modulus of the local hexatic order. 
        The c.o.m. trajectory is shown in black and the local red arrows 
        the active force coarse-grained on a square grid.
        (d) Mass evolution of two active clusters with different initial masses
        (lower curves) compared to a typical  passive one formed by attractive particles (upper red curve).
        The horizontal dashed lines are the time averages.
        (e) Probability distribution of the instantaneous masses.  The shaded gray
        region
        indicates the clusters used to calculate $D(M)$.
        (f) Mass dependence of $D$ for ABP clusters (diamonds Pe = 80, squares Pe = 100, triangles Pe = 200)
        and clusters of passive attractive particles
        (red bullets, translated upwards by a factor 40). In the ABP case, we identify two algebraic regimes, $D\sim M^{-\alpha}$, with  fitted
        $\alpha \sim 0.5$ (solid line) and $\alpha = 1$ (dashed line), for small and large masses.
        In the passive case, the slope $\alpha = 1$  is  very close to the data.
        The temperature is $T=0.05$ (active) and
        $T=0.3$ (passive).
    }
    \label{fig:tres}
\end{figure}

In order to follow the clusters  over longer periods,
we placed  individual ones in a box with linear size of the order of the inter-cluster distance in the bulk, and we 
filled space with
an active gas mimicking the global conditions of the
interacting clusters. The three snapshots in Fig.~\ref{fig:tres}(a)-(c)
illustrate the evolution of one such cluster. The
light and rather straight lines inside the cluster
are at interfaces between different orientational (hexatic) orders.
The cluster changes form, by opening up gas bubbles~\cite{Tjhung18,Caporusso20,Shi20}
and growing protrusions,
and displaces by a kind of crawling mechanism.
Movie M1 in SM displays the  evolution of a similar cluster.
Sudden rearrangements yield rather large mass (and shape) variations, see Fig.~\ref{fig:tres}(d).
The active behavior is confronted to the passive one
in the upper (red) curve which is smoother.
The pdf of the instantaneous masses is plotted in
Fig.~\ref{fig:tres}(e).

Figure~\ref{fig:tres}(f) represents the diffusion coefficient $D$ {\it vs.} the 
cluster mass $M$, $D$ is
the average over the instantaneous values in the gray window of Fig.~\ref{fig:tres}(e). 
Differently from the bulk data in Fig.~\ref{fig:dos}(b),
$D$ displays two  regimes, described by $D\sim M^{-0.5}$ and $M^{-1}$.
Clusters of passive attractive particles with similar packing fractions 
exhibit $D\sim M^{-1}$ over the full mass range (red bullets~\cite{SM}).

The large mass  dependence can be estimated with a simple argument,
after characterising  the forces exerted on an isolated cluster.
The net potential force vanishes due to the action-reaction principle,
and the total torques are negligible, Fig.~\ref{fig:quattro}(d),
supporting the fact that clusters do not rotate significantly.
The total active force, $\bm{F}= F_{\rm act} \sum_{i=1}^{M/m}  {\bf n}_i$, 
has exponentially decaying temporal
correlations, Fig.~\ref{fig:quattro}(a), with characteristic time $\approx\tau_p$,
and, at equal times,
$\langle \bm{F}^2(0)\rangle \propto (M/m) F_{\rm act}^2$,
Fig.~\ref{fig:quattro}(b), 
for all masses studied.
The calculation in Sec.~D of the SM, which treats the active gas as a white noise bath,
and the active forces as independent, yields
$D 
\propto M^{-1}$.

However, two mechanisms conspire against $D \propto M^{-1}$
 for small enough clusters.
One concerns the clusters themselves. Firstly, the pdf of the angle $\varphi$ between   ${\bf n}_i$
and  $\bm{F}$
   has  extra weight around $\varphi =0$  (more pronounced for smaller clusters), Fig.~\ref{fig:quattro}(c). Indeed, 
   the active forces on boundary particles can be more aligned
with  the direction of motion, Fig.~\ref{fig:quattro}(d),  
and make the small clusters more mobile.  
Secondly, the cluster's motion  exhibits sudden events
in which pieces  detach and glue back after coordinated collisions (Fig.~\ref{fig:tres}(a)-(c) and M1). 
Thirdly, it is almost impossible to find large isolated clusters in the scaling regime at not too low packing fraction 
(and  their form is much more contorted than in isolation, see Fig.~\ref{fig:uno}(c) and below). 
The impact of other clusters, favored by their own mobility,
cannot be neglected.
The other mechanism focuses on the effects of the active gas around the clusters. The derivation
in~\cite{Solon22} of the size dependence of the diffusion coefficient of a spherical passive tracer immersed in an active 
gas yields $D\sim M^{-1/2}$ as  in  Fig.~\ref{fig:tres}(f), with the cross-over size increasing with  the persistence length
of the gas particles. 

\begin{figure}[t!]
    \centering
    \includegraphics[width=8.8cm]{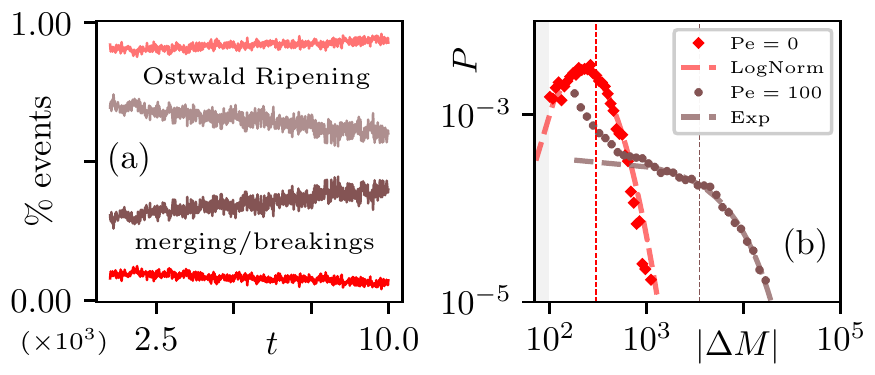}
    \vspace{-0.35cm}
    \caption{{\bf Collisions.}
            (a) Percentages of events with mass changes $|\Delta M| < 100$ (above), and 
           those with  $|\Delta M| \geq 100$ (below) for Pe = 0 (red), 100  (brown). 
            (b) Distributions of absolute mass changes at $t=10^3 >t_s$ for $\delta t = 10$. 
            The dashed lines are fits with the functions  in the key.
            }
    \label{fig:quattro-new}
\end{figure}

We now tailor the tracking algorithm to measure mass variations~\cite{SM}, 
and distinguish Ostwald ripening from cluster-cluster aggregation
as cluster growth mechanisms in the bulk.
In Fig.~\ref{fig:quattro-new}(a) we show
the normalised counting of small mass variation events, 
that we attribute to evaporation/condensation of particles
(above), and large mass variation events due to breaking/merging between clusters 
(below), during the observed timeframe $\delta t=10$~\cite{SM}.
The threshold is set at the cluster mass variation 
$|\Delta M| = |M_{t+\delta t}-M_t| =100$ but similar results are found for other 
choices nearby.
In the passive case (red data) the former percentage is very close to one and slowly grows in time, 
underscoring that Ostwald ripening is the dominant mechanism at long timescales~\cite{Jack16}. 
Instead, for ABPs, the trend is the opposite, with  the  percentage of large cluster rearrangements
due to aggregation and/or break up becoming more and more important at long times. 
Further differences are shown by the distributions of  $|\Delta M|$
at the beginning of the scaling regime, 
plotted in Fig.~\ref{fig:quattro-new}(b).
The passive system displays a clear peak,  due to the aggregation of clusters with a preferred averaged size, 
see Fig.~\ref{fig:dist-Rg-attractive}(b), well fitted by a log-normal function.
Instead, the form in the active system is rather different
and can be attributed to merging of approximately independent  clusters with exponential size distribution, 
see Fig.~\ref{fig:dist-m-and-dm-1000}. 

\begin{figure}[b!]
    \centering
    \hspace{1.5cm}
    \includegraphics[width=8cm]{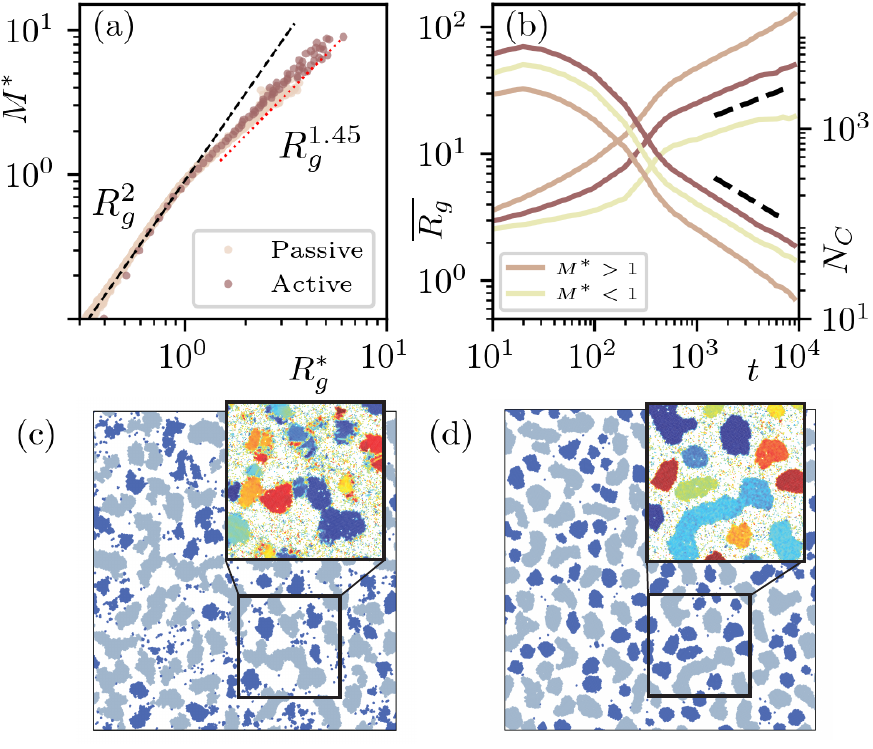}
    \caption{{\bf Clusters geometry.}
        (a) Scatter plot of $M^*=M/\overline M$ against $R_g^*=R_g/\overline R_g$,
        for passive and active (Pe = 100) clusters.
        (b) Left scale:  the averaged radius of gyration
        of regular ($M^*<1$), fractal ($M^*>1$) and all clusters.
        Right scale:  the number of clusters.
        The dashed lines are fits of the total averaged curves, 
 $t^{0.31}$ and $t^{-0.65}$. Separate fits of the fractal and regular $\overline R_g$ data yield $z$,
            close to $2.2$ and $3.85$, respectively. 
        A snapshot of (c) the ABP system and (d) the passive system in the scaling regime.
        Fractal clusters are gray and regular ones blue. The zooms in the boxes 
        represent the local hexatic order with a color map.
    }
    \label{fig:five}
\end{figure}

Not all clusters have the same shape 
and their geometry may influence the growth of the dense component:
we now analyze this possibility.
At all times, regular (with fractal dimension $d_f$ equal to the dimension
of space $d=2$) and fractal (with $d_f<2$) clusters co-exist. In  Fig.~\ref{fig:five}(a)
we plot $M^*$,  the mass normalized by its average $\overline M$, against $R_g^*$,  the radius of gyration
also normalized by its average $\overline R_g$,
of individual clusters sampled in the scaling regime $t = 10^3 - 10^5$.
The data
follow $M^* \sim {R^*_g}^{d_f}$, with $d_f\simeq 2$ for $M^*<1$,
and $d_f\simeq 1.45$ for $M^*>1$
both in the active and the passive case.
The crossover at $M^* \sim R_g^* \sim 1$
is  controlled by $\overline M$ and $\overline R_g$, which depend on time in the form reported in
Fig.~\ref{fig:uno}(a) and~\ref{fig:five}(b), respectively.  This is similar to what observed in lattice spin models~\cite{Bray}.
For diffusion limited cluster-cluster aggregation $d_f \simeq 1.45$  in
$d = 2$~\cite{Kolb83,Meakin83}, as well as  in passive Lennard-Jones systems at very low $\phi$~\cite{Paul21}.
Having identified two kinds of clusters we calculated the average mass and radii separately. In Fig.~\ref{fig:five}(b) we show
the growth of the mean $R_g$ for regular (lower curve) and fractal (upper) clusters,
as well as the completely averaged
$\overline R_g$  (middle).
All kinds of clusters enter the scaling regime at the same $t_s$.
Fractal (regular) clusters grow slightly faster (slower) than the average within our numerical precision.
There are roughly 2-3 times more regular than fractal clusters, see the $N_C$ curves on the right scale.


Having found clusters' diffusion and aggregation in the active system,  we are tempted to test
the Smoluchowski relation  $z = d_f\,(1+\alpha)$, valid  for Brownian diffusive
binary cluster aggregation under mean-field assumptions~\cite{Jullien92}. For $M<\overline M$ we found
$d_f\approx 2$ and $\alpha\approx 0.5$, leading to $z\approx 3$.
Instead, for $M>\overline M$ while $d_f$ changes to $d_f\approx 1.45$, we still measure $\alpha \approx 0.5$
in the bulk, and this gives $z\approx 2.2$ which is very close to the value fitted in Fig.~\ref{fig:five}(b).
Had we claimed that $\alpha$ should become $1$ for sufficiently large masses, as measured for the 
clusters extracted from the bulk, then $z\approx 2.9$, very close to $3$.
For the equilibrium system, instead, cluster aggregation becomes irrelevant at long times (movie M3 and Fig.~\ref{fig:quattro-new}) 
and there is no reason to use the Smoluchowski relation.

Our analysis evidences a number of important differences between MIPS and conservative phase separation.
Active clusters
are made of a patchwork of different hexatic orders (with dynamic bubbles in cavitation~\cite{Tjhung18,Caporusso20,Shi20}), while passive ones are homogeneous.
As appreciated in M2 and M3 \cite{SM}, and confirmed in Fig.~\ref{fig:quattro-new},
passive small clusters evaporate (\`a la Ostwald ripening)
while active ones can break and sometimes recombine (\`a la Smoluchowski).
The interfaces between the different (hexatic) orientations of the colliding ABP clusters do not heal.
The external surfaces of the active clusters are much rougher than the ones of the conservative clusters,
see Fig.~\ref{fig:five}(c)-(d). Due to the enhanced cluster's diffusion constant, $D \sim {\rm Pe}^2/M^{0.5}$,
growth is faster for active than passive particles.
In both cases there is co-existence of small regular  and large fractal  clusters with the
crossover at the averaged scales.
We conclude that, although the global growth laws in MIPS and equilibrium phase separation are both $t^{1/3}$,
a mesoscopic  investigation demonstrates that the dynamics of MIPS clusters is quite different from equilibrium demixing droplets.
In MIPS, simple Ostwald ripening combines with fragmentation and aggregation of diffusive clusters to yield  the familiar
$t^{1/3}$ growth in a highly non-trivial manner.


\vspace{0.25cm}

\noindent
{\it Acknowledgments.}
We used the computers Lenovo NeXt\-Scale MARCONI at CINECA (Project  INF16-field\-turb) under  CINECA-INFN  agreement,
and MareNostrum  at the BSC.
Research supported by the MIUR project PRIN/2020 PFCXPE
``Response, control and learning: building new manipulation strategies in
living and engineered active matter"
(CC, GG, AS), JIN RTI2018-099032-J-I00  (MCI/AEI/FEDER, UE) (DL) and
ANR 226573 THEMA (LFC). We thank R. Jack, A. Solon and F. van Wijland
for very useful suggestions.

\let\oldaddcontentsline\addcontentsline
\renewcommand{\addcontentsline}[3]{}

\bibliographystyle{apsrev4-1}
\bibliography{ABP_clusters_nov22.bib}

\begin{thebibliography}{55}%
\makeatletter
\providecommand \@ifxundefined [1]{%
 \@ifx{#1\undefined}
}%
\providecommand \@ifnum [1]{%
 \ifnum #1\expandafter \@firstoftwo
 \else \expandafter \@secondoftwo
 \fi
}%
\providecommand \@ifx [1]{%
 \ifx #1\expandafter \@firstoftwo
 \else \expandafter \@secondoftwo
 \fi
}%
\providecommand \natexlab [1]{#1}%
\providecommand \enquote  [1]{``#1''}%
\providecommand \bibnamefont  [1]{#1}%
\providecommand \bibfnamefont [1]{#1}%
\providecommand \citenamefont [1]{#1}%
\providecommand \href@noop [0]{\@secondoftwo}%
\providecommand \href [0]{\begingroup \@sanitize@url \@href}%
\providecommand \@href[1]{\@@startlink{#1}\@@href}%
\providecommand \@@href[1]{\endgroup#1\@@endlink}%
\providecommand \@sanitize@url [0]{\catcode `\\12\catcode `\$12\catcode
  `\&12\catcode `\#12\catcode `\^12\catcode `\_12\catcode `\%12\relax}%
\providecommand \@@startlink[1]{}%
\providecommand \@@endlink[0]{}%
\providecommand \url  [0]{\begingroup\@sanitize@url \@url }%
\providecommand \@url [1]{\endgroup\@href {#1}{\urlprefix }}%
\providecommand \urlprefix  [0]{URL }%
\providecommand \Eprint [0]{\href }%
\providecommand \doibase [0]{http://dx.doi.org/}%
\providecommand \selectlanguage [0]{\@gobble}%
\providecommand \bibinfo  [0]{\@secondoftwo}%
\providecommand \bibfield  [0]{\@secondoftwo}%
\providecommand \translation [1]{[#1]}%
\providecommand \BibitemOpen [0]{}%
\providecommand \bibitemStop [0]{}%
\providecommand \bibitemNoStop [0]{.\EOS\space}%
\providecommand \EOS [0]{\spacefactor3000\relax}%
\providecommand \BibitemShut  [1]{\csname bibitem#1\endcsname}%
\let\auto@bib@innerbib\@empty
\bibitem [{\citenamefont {Lifshitz}\ and\ \citenamefont
  {Slyozov}(1961)}]{lifshitz1961kinetics}%
  \BibitemOpen
  \bibfield  {author} {\bibinfo {author} {\bibfnamefont {I.~M.}\ \bibnamefont
  {Lifshitz}}\ and\ \bibinfo {author} {\bibfnamefont {V.~V.}\ \bibnamefont
  {Slyozov}},\ }\href@noop {} {\bibfield  {journal} {\bibinfo  {journal} {J.
  Phys. Chem. Solids}\ }\textbf {\bibinfo {volume} {19}},\ \bibinfo {pages}
  {35} (\bibinfo {year} {1961})}\BibitemShut {NoStop}%
\bibitem [{\citenamefont {Wagner}(1961)}]{wagner1961theory}%
  \BibitemOpen
  \bibfield  {author} {\bibinfo {author} {\bibfnamefont {C.}~\bibnamefont
  {Wagner}},\ }\href@noop {} {\bibfield  {journal} {\bibinfo  {journal} {Z.
  Elektrochem.}\ }\textbf {\bibinfo {volume} {65}},\ \bibinfo {pages} {581}
  (\bibinfo {year} {1961})}\BibitemShut {NoStop}%
\bibitem [{\citenamefont {Voorhees}(1985)}]{voorhees1985theory}%
  \BibitemOpen
  \bibfield  {author} {\bibinfo {author} {\bibfnamefont {P.~W.}\ \bibnamefont
  {Voorhees}},\ }\href@noop {} {\bibfield  {journal} {\bibinfo  {journal} {J.
  Stat. Phys.}\ }\textbf {\bibinfo {volume} {38}},\ \bibinfo {pages} {231}
  (\bibinfo {year} {1985})}\BibitemShut {NoStop}%
\bibitem [{\citenamefont {Hohenberg}\ and\ \citenamefont
  {Halperin}(1977)}]{HohenbergRev}%
  \BibitemOpen
  \bibfield  {author} {\bibinfo {author} {\bibfnamefont {P.~C.}\ \bibnamefont
  {Hohenberg}}\ and\ \bibinfo {author} {\bibfnamefont {B.~I.}\ \bibnamefont
  {Halperin}},\ }\href@noop {} {\bibfield  {journal} {\bibinfo  {journal} {Rev.
  Mod. Phys.}\ }\textbf {\bibinfo {volume} {49}},\ \bibinfo {pages} {435}
  (\bibinfo {year} {1977})}\BibitemShut {NoStop}%
\bibitem [{\citenamefont {Onuki}(2002)}]{OnukiBook}%
  \BibitemOpen
  \bibfield  {author} {\bibinfo {author} {\bibfnamefont {A.}~\bibnamefont
  {Onuki}},\ }\href@noop {} {\emph {\bibinfo {title} {Phase transition
  dynamics}}}\ (\bibinfo  {publisher} {Cambridge University Press},\ \bibinfo
  {year} {2002})\BibitemShut {NoStop}%
\bibitem [{\citenamefont {Bray}(2002)}]{BrayRev}%
  \BibitemOpen
  \bibfield  {author} {\bibinfo {author} {\bibfnamefont {A.~J.}\ \bibnamefont
  {Bray}},\ }\href@noop {} {\bibfield  {journal} {\bibinfo  {journal} {Adv. in
  Phys.}\ }\textbf {\bibinfo {volume} {51}},\ \bibinfo {pages} {481} (\bibinfo
  {year} {2002})}\BibitemShut {NoStop}%
\bibitem [{\citenamefont {Bechinger}\ \emph {et~al.}(2016)\citenamefont
  {Bechinger}, \citenamefont {Di~Leonardo}, \citenamefont {L{\"o}wen},
  \citenamefont {Reichhardt},\ and\ \citenamefont {Volpe}}]{BechingerRev}%
  \BibitemOpen
  \bibfield  {author} {\bibinfo {author} {\bibfnamefont {C.}~\bibnamefont
  {Bechinger}}, \bibinfo {author} {\bibfnamefont {R.}~\bibnamefont
  {Di~Leonardo}}, \bibinfo {author} {\bibfnamefont {H.}~\bibnamefont
  {L{\"o}wen}}, \bibinfo {author} {\bibfnamefont {C.}~\bibnamefont
  {Reichhardt}}, \ and\ \bibinfo {author} {\bibfnamefont {G.}~\bibnamefont
  {Volpe}},\ }\href@noop {} {\bibfield  {journal} {\bibinfo  {journal} {Rev.
  Mod. Phys.}\ }\textbf {\bibinfo {volume} {88}},\ \bibinfo {pages} {045006}
  (\bibinfo {year} {2016})}\BibitemShut {NoStop}%
\bibitem [{\citenamefont {Shaebani}\ \emph {et~al.}(2020)\citenamefont
  {Shaebani}, \citenamefont {Wysocki}, \citenamefont {Winkler}, \citenamefont
  {Gompper},\ and\ \citenamefont {Rieger}}]{WinklerRev}%
  \BibitemOpen
  \bibfield  {author} {\bibinfo {author} {\bibfnamefont {M.~R.}\ \bibnamefont
  {Shaebani}}, \bibinfo {author} {\bibfnamefont {A.}~\bibnamefont {Wysocki}},
  \bibinfo {author} {\bibfnamefont {R.~G.}\ \bibnamefont {Winkler}}, \bibinfo
  {author} {\bibfnamefont {G.}~\bibnamefont {Gompper}}, \ and\ \bibinfo
  {author} {\bibfnamefont {H.}~\bibnamefont {Rieger}},\ }\href@noop {}
  {\bibfield  {journal} {\bibinfo  {journal} {Nat. Rev. Phys.}\ }\textbf
  {\bibinfo {volume} {2}},\ \bibinfo {pages} {181} (\bibinfo {year}
  {2020})}\BibitemShut {NoStop}%
\bibitem [{\citenamefont {O’Byrne}\ \emph {et~al.}(2022)\citenamefont
  {O’Byrne}, \citenamefont {Kafri}, \citenamefont {Tailleur},\ and\
  \citenamefont {van Wijland}}]{TailleurRev}%
  \BibitemOpen
  \bibfield  {author} {\bibinfo {author} {\bibfnamefont {J.}~\bibnamefont
  {O’Byrne}}, \bibinfo {author} {\bibfnamefont {Y.}~\bibnamefont {Kafri}},
  \bibinfo {author} {\bibfnamefont {J.}~\bibnamefont {Tailleur}}, \ and\
  \bibinfo {author} {\bibfnamefont {F.}~\bibnamefont {van Wijland}},\
  }\href@noop {} {\bibfield  {journal} {\bibinfo  {journal} {Nat. Rev. Phys.}\
  }\textbf {\bibinfo {volume} {4}},\ \bibinfo {pages} {167} (\bibinfo {year}
  {2022})}\BibitemShut {NoStop}%
\bibitem [{\citenamefont {Tailleur}\ and\ \citenamefont
  {Cates}(2008)}]{TailleurCates2008}%
  \BibitemOpen
  \bibfield  {author} {\bibinfo {author} {\bibfnamefont {J.}~\bibnamefont
  {Tailleur}}\ and\ \bibinfo {author} {\bibfnamefont {M.~E.}\ \bibnamefont
  {Cates}},\ }\href@noop {} {\bibfield  {journal} {\bibinfo  {journal} {Phys.
  Rev. Lett.}\ }\textbf {\bibinfo {volume} {100}},\ \bibinfo {pages} {218103}
  (\bibinfo {year} {2008})}\BibitemShut {NoStop}%
\bibitem [{\citenamefont {Cates}\ and\ \citenamefont
  {Tailleur}(2015)}]{CatesRev}%
  \BibitemOpen
  \bibfield  {author} {\bibinfo {author} {\bibfnamefont {M.~E.}\ \bibnamefont
  {Cates}}\ and\ \bibinfo {author} {\bibfnamefont {J.}~\bibnamefont
  {Tailleur}},\ }\href@noop {} {\bibfield  {journal} {\bibinfo  {journal}
  {Annu. Rev. Cond. Matt. Phys.}\ }\textbf {\bibinfo {volume} {6}},\ \bibinfo
  {pages} {219} (\bibinfo {year} {2015})}\BibitemShut {NoStop}%
\bibitem [{\citenamefont {Romanczuk}\ \emph {et~al.}(2012)\citenamefont
  {Romanczuk}, \citenamefont {B{\"a}r}, \citenamefont {Ebeling}, \citenamefont
  {Lindner},\ and\ \citenamefont {Schimansky-Geier}}]{romanczuk2012active}%
  \BibitemOpen
  \bibfield  {author} {\bibinfo {author} {\bibfnamefont {P.}~\bibnamefont
  {Romanczuk}}, \bibinfo {author} {\bibfnamefont {M.}~\bibnamefont {B{\"a}r}},
  \bibinfo {author} {\bibfnamefont {W.}~\bibnamefont {Ebeling}}, \bibinfo
  {author} {\bibfnamefont {B.}~\bibnamefont {Lindner}}, \ and\ \bibinfo
  {author} {\bibfnamefont {L.}~\bibnamefont {Schimansky-Geier}},\ }\href@noop
  {} {\bibfield  {journal} {\bibinfo  {journal} {Eur. Phys. J. Sp. Topics}\
  }\textbf {\bibinfo {volume} {202}},\ \bibinfo {pages} {1} (\bibinfo {year}
  {2012})}\BibitemShut {NoStop}%
\bibitem [{\citenamefont {Fily}\ and\ \citenamefont
  {Marchetti}(2012)}]{Fily2012}%
  \BibitemOpen
  \bibfield  {author} {\bibinfo {author} {\bibfnamefont {Y.}~\bibnamefont
  {Fily}}\ and\ \bibinfo {author} {\bibfnamefont {M.~C.}\ \bibnamefont
  {Marchetti}},\ }\href@noop {} {\bibfield  {journal} {\bibinfo  {journal}
  {Phys. Rev. Lett.}\ }\textbf {\bibinfo {volume} {108}},\ \bibinfo {pages}
  {235702} (\bibinfo {year} {2012})}\BibitemShut {NoStop}%
\bibitem [{\citenamefont {Redner}\ \emph {et~al.}(2013)\citenamefont {Redner},
  \citenamefont {Hagan},\ and\ \citenamefont {Baskaran}}]{Redner2013f}%
  \BibitemOpen
  \bibfield  {author} {\bibinfo {author} {\bibfnamefont {G.~S.}\ \bibnamefont
  {Redner}}, \bibinfo {author} {\bibfnamefont {M.~F.}\ \bibnamefont {Hagan}}, \
  and\ \bibinfo {author} {\bibfnamefont {A.}~\bibnamefont {Baskaran}},\
  }\href@noop {} {\bibfield  {journal} {\bibinfo  {journal} {Phys. Rev. Lett.}\
  }\textbf {\bibinfo {volume} {110}},\ \bibinfo {pages} {055701} (\bibinfo
  {year} {2013})}\BibitemShut {NoStop}%
\bibitem [{\citenamefont {Stenhammar}\ \emph {et~al.}(2014)\citenamefont
  {Stenhammar}, \citenamefont {Marenduzzo}, \citenamefont {Allen},\ and\
  \citenamefont {Cates}}]{Stenhammar2014}%
  \BibitemOpen
  \bibfield  {author} {\bibinfo {author} {\bibfnamefont {J.}~\bibnamefont
  {Stenhammar}}, \bibinfo {author} {\bibfnamefont {D.}~\bibnamefont
  {Marenduzzo}}, \bibinfo {author} {\bibfnamefont {R.~J.}\ \bibnamefont
  {Allen}}, \ and\ \bibinfo {author} {\bibfnamefont {M.~E.}\ \bibnamefont
  {Cates}},\ }\href@noop {} {\bibfield  {journal} {\bibinfo  {journal} {Soft
  Matter}\ }\textbf {\bibinfo {volume} {10}},\ \bibinfo {pages} {1489}
  (\bibinfo {year} {2014})}\BibitemShut {NoStop}%
\bibitem [{\citenamefont {Wysocki}\ \emph {et~al.}(2014)\citenamefont
  {Wysocki}, \citenamefont {Winkler},\ and\ \citenamefont {Gompper}}]{Wysocki}%
  \BibitemOpen
  \bibfield  {author} {\bibinfo {author} {\bibfnamefont {A.}~\bibnamefont
  {Wysocki}}, \bibinfo {author} {\bibfnamefont {R.~G.}\ \bibnamefont
  {Winkler}}, \ and\ \bibinfo {author} {\bibfnamefont {G.}~\bibnamefont
  {Gompper}},\ }\href@noop {} {\bibfield  {journal} {\bibinfo  {journal} {EPL}\
  }\textbf {\bibinfo {volume} {105}},\ \bibinfo {pages} {48004} (\bibinfo
  {year} {2014})}\BibitemShut {NoStop}%
\bibitem [{\citenamefont {Caporusso}\ \emph {et~al.}(2020)\citenamefont
  {Caporusso}, \citenamefont {Digregorio}, \citenamefont {Levis}, \citenamefont
  {Cugliandolo},\ and\ \citenamefont {Gonnella}}]{Caporusso20}%
  \BibitemOpen
  \bibfield  {author} {\bibinfo {author} {\bibfnamefont {C.~B.}\ \bibnamefont
  {Caporusso}}, \bibinfo {author} {\bibfnamefont {L.}~\bibnamefont
  {Digregorio}}, \bibinfo {author} {\bibfnamefont {D.}~\bibnamefont {Levis}},
  \bibinfo {author} {\bibfnamefont {L.~F.}\ \bibnamefont {Cugliandolo}}, \ and\
  \bibinfo {author} {\bibfnamefont {G.}~\bibnamefont {Gonnella}},\ }\href@noop
  {} {\bibfield  {journal} {\bibinfo  {journal} {Phys. Rev. Lett.}\ }\textbf
  {\bibinfo {volume} {125}},\ \bibinfo {pages} {178004} (\bibinfo {year}
  {2020})}\BibitemShut {NoStop}%
\bibitem [{\citenamefont {Digregorio}\ \emph {et~al.}(2018)\citenamefont
  {Digregorio}, \citenamefont {Levis}, \citenamefont {Suma}, \citenamefont
  {Cugliandolo}, \citenamefont {Gonnella},\ and\ \citenamefont
  {Pagonabarraga}}]{PRLino}%
  \BibitemOpen
  \bibfield  {author} {\bibinfo {author} {\bibfnamefont {P.}~\bibnamefont
  {Digregorio}}, \bibinfo {author} {\bibfnamefont {D.}~\bibnamefont {Levis}},
  \bibinfo {author} {\bibfnamefont {A.}~\bibnamefont {Suma}}, \bibinfo {author}
  {\bibfnamefont {L.~F.}\ \bibnamefont {Cugliandolo}}, \bibinfo {author}
  {\bibfnamefont {G.}~\bibnamefont {Gonnella}}, \ and\ \bibinfo {author}
  {\bibfnamefont {I.}~\bibnamefont {Pagonabarraga}},\ }\href@noop {} {\bibfield
   {journal} {\bibinfo  {journal} {Phys. Rev. Lett.}\ }\textbf {\bibinfo
  {volume} {121}},\ \bibinfo {pages} {098003} (\bibinfo {year}
  {2018})}\BibitemShut {NoStop}%
\bibitem [{\citenamefont {Caprini}\ \emph {et~al.}(2020)\citenamefont
  {Caprini}, \citenamefont {Marconi},\ and\ \citenamefont
  {Puglisi}}]{Caprini20}%
  \BibitemOpen
  \bibfield  {author} {\bibinfo {author} {\bibfnamefont {L.}~\bibnamefont
  {Caprini}}, \bibinfo {author} {\bibfnamefont {U.~M.~B.}\ \bibnamefont
  {Marconi}}, \ and\ \bibinfo {author} {\bibfnamefont {A.}~\bibnamefont
  {Puglisi}},\ }\href@noop {} {\bibfield  {journal} {\bibinfo  {journal} {Phys.
  Rev. Lett.}\ }\textbf {\bibinfo {volume} {124}},\ \bibinfo {pages} {078001}
  (\bibinfo {year} {2020})}\BibitemShut {NoStop}%
\bibitem [{\citenamefont {Maggi}\ \emph {et~al.}(2021)\citenamefont {Maggi},
  \citenamefont {Paoluzzi}, \citenamefont {Crisanti}, \citenamefont
  {Zaccarelli},\ and\ \citenamefont {Gnan}}]{Gnan21}%
  \BibitemOpen
  \bibfield  {author} {\bibinfo {author} {\bibfnamefont {C.}~\bibnamefont
  {Maggi}}, \bibinfo {author} {\bibfnamefont {M.}~\bibnamefont {Paoluzzi}},
  \bibinfo {author} {\bibfnamefont {A.}~\bibnamefont {Crisanti}}, \bibinfo
  {author} {\bibfnamefont {E.}~\bibnamefont {Zaccarelli}}, \ and\ \bibinfo
  {author} {\bibfnamefont {N.}~\bibnamefont {Gnan}},\ }\href@noop {} {\bibfield
   {journal} {\bibinfo  {journal} {Soft Matter}\ }\textbf {\bibinfo {volume}
  {17}},\ \bibinfo {pages} {3807} (\bibinfo {year} {2021})}\BibitemShut
  {NoStop}%
\bibitem [{\citenamefont {Mart\'{\i}n-Roca}\ \emph {et~al.}(2021)\citenamefont
  {Mart\'{\i}n-Roca}, \citenamefont {Mart{\'{\i}}nez}, \citenamefont
  {Alexander}, \citenamefont {Diez}, \citenamefont {Aarts}, \citenamefont
  {Alarc{\'o}n}, \citenamefont {Ram\'{\i}rez},\ and\ \citenamefont
  {Valeriani}}]{Valeriani21}%
  \BibitemOpen
  \bibfield  {author} {\bibinfo {author} {\bibfnamefont {J.}~\bibnamefont
  {Mart\'{\i}n-Roca}}, \bibinfo {author} {\bibfnamefont {R.}~\bibnamefont
  {Mart{\'{\i}}nez}}, \bibinfo {author} {\bibfnamefont {L.~C.}\ \bibnamefont
  {Alexander}}, \bibinfo {author} {\bibfnamefont {A.~L.}\ \bibnamefont {Diez}},
  \bibinfo {author} {\bibfnamefont {D.~G. A.~L.}\ \bibnamefont {Aarts}},
  \bibinfo {author} {\bibfnamefont {F.}~\bibnamefont {Alarc{\'o}n}}, \bibinfo
  {author} {\bibfnamefont {J.}~\bibnamefont {Ram\'{\i}rez}}, \ and\ \bibinfo
  {author} {\bibfnamefont {C.}~\bibnamefont {Valeriani}},\ }\href@noop {}
  {\bibfield  {journal} {\bibinfo  {journal} {J. Chem. Phys.}\ }\textbf
  {\bibinfo {volume} {154}},\ \bibinfo {pages} {164901} (\bibinfo {year}
  {2021})}\BibitemShut {NoStop}%
\bibitem [{\citenamefont {Yang}\ \emph {et~al.}(2022)\citenamefont {Yang},
  \citenamefont {Ni},\ and\ \citenamefont {{Pica~Ciamarra}}}]{PicaCiamarra22}%
  \BibitemOpen
  \bibfield  {author} {\bibinfo {author} {\bibfnamefont {J.}~\bibnamefont
  {Yang}}, \bibinfo {author} {\bibfnamefont {R.}~\bibnamefont {Ni}}, \ and\
  \bibinfo {author} {\bibfnamefont {M.}~\bibnamefont {{Pica~Ciamarra}}},\
  }\href@noop {} {\bibfield  {journal} {\bibinfo  {journal} {Phys. Rev. E}\
  }\textbf {\bibinfo {volume} {106}},\ \bibinfo {pages} {L012601} (\bibinfo
  {year} {2022})}\BibitemShut {NoStop}%
\bibitem [{\citenamefont {Cates}(2019)}]{cates2019active}%
  \BibitemOpen
  \bibfield  {author} {\bibinfo {author} {\bibfnamefont {M.~E.}\ \bibnamefont
  {Cates}},\ }\href@noop {} {\bibfield  {journal} {\bibinfo  {journal} {arXiv
  preprint arXiv:1904.01330}\ } (\bibinfo {year} {2019})}\BibitemShut {NoStop}%
\bibitem [{\citenamefont {Stenhammar}\ \emph {et~al.}(2013)\citenamefont
  {Stenhammar}, \citenamefont {Tiribocchi}, \citenamefont {Allen},
  \citenamefont {Marenduzzo},\ and\ \citenamefont {Cates}}]{Stenhammar13}%
  \BibitemOpen
  \bibfield  {author} {\bibinfo {author} {\bibfnamefont {J.}~\bibnamefont
  {Stenhammar}}, \bibinfo {author} {\bibfnamefont {A.}~\bibnamefont
  {Tiribocchi}}, \bibinfo {author} {\bibfnamefont {R.~J.}\ \bibnamefont
  {Allen}}, \bibinfo {author} {\bibfnamefont {D.}~\bibnamefont {Marenduzzo}}, \
  and\ \bibinfo {author} {\bibfnamefont {M.~E.}\ \bibnamefont {Cates}},\
  }\href@noop {} {\bibfield  {journal} {\bibinfo  {journal} {Phys. Rev. Lett.}\
  }\textbf {\bibinfo {volume} {111}},\ \bibinfo {pages} {145702} (\bibinfo
  {year} {2013})}\BibitemShut {NoStop}%
\bibitem [{\citenamefont {Wittkowski}\ \emph {et~al.}(2014)\citenamefont
  {Wittkowski}, \citenamefont {Tiribocchi}, \citenamefont {Stenhammar},
  \citenamefont {Allen}, \citenamefont {Marenduzzo},\ and\ \citenamefont
  {Cates}}]{wittkowski2014}%
  \BibitemOpen
  \bibfield  {author} {\bibinfo {author} {\bibfnamefont {R.}~\bibnamefont
  {Wittkowski}}, \bibinfo {author} {\bibfnamefont {A.}~\bibnamefont
  {Tiribocchi}}, \bibinfo {author} {\bibfnamefont {J.}~\bibnamefont
  {Stenhammar}}, \bibinfo {author} {\bibfnamefont {R.~J.}\ \bibnamefont
  {Allen}}, \bibinfo {author} {\bibfnamefont {D.}~\bibnamefont {Marenduzzo}}, \
  and\ \bibinfo {author} {\bibfnamefont {M.~E.}\ \bibnamefont {Cates}},\
  }\href@noop {} {\bibfield  {journal} {\bibinfo  {journal} {Nature Comm.}\
  }\textbf {\bibinfo {volume} {5}},\ \bibinfo {pages} {1} (\bibinfo {year}
  {2014})}\BibitemShut {NoStop}%
\bibitem [{\citenamefont {Speck}\ \emph {et~al.}(2014)\citenamefont {Speck},
  \citenamefont {Bialk\'e}, \citenamefont {Menzel},\ and\ \citenamefont
  {L\"owen}}]{Speck14}%
  \BibitemOpen
  \bibfield  {author} {\bibinfo {author} {\bibfnamefont {T.}~\bibnamefont
  {Speck}}, \bibinfo {author} {\bibfnamefont {J.}~\bibnamefont {Bialk\'e}},
  \bibinfo {author} {\bibfnamefont {A.~M.}\ \bibnamefont {Menzel}}, \ and\
  \bibinfo {author} {\bibfnamefont {H.}~\bibnamefont {L\"owen}},\ }\href@noop
  {} {\bibfield  {journal} {\bibinfo  {journal} {Phys. Rev. Lett.}\ }\textbf
  {\bibinfo {volume} {112}},\ \bibinfo {pages} {218304} (\bibinfo {year}
  {2014})}\BibitemShut {NoStop}%
\bibitem [{\citenamefont {Speck}\ \emph {et~al.}(2015)\citenamefont {Speck},
  \citenamefont {Menzel}, \citenamefont {Bialk\'e},\ and\ \citenamefont
  {L\"owen}}]{Speck15}%
  \BibitemOpen
  \bibfield  {author} {\bibinfo {author} {\bibfnamefont {T.}~\bibnamefont
  {Speck}}, \bibinfo {author} {\bibfnamefont {A.~M.}\ \bibnamefont {Menzel}},
  \bibinfo {author} {\bibfnamefont {J.}~\bibnamefont {Bialk\'e}}, \ and\
  \bibinfo {author} {\bibfnamefont {H.}~\bibnamefont {L\"owen}},\ }\href@noop
  {} {\bibfield  {journal} {\bibinfo  {journal} {J. Chem. Phys.}\ }\textbf
  {\bibinfo {volume} {142}},\ \bibinfo {pages} {224109} (\bibinfo {year}
  {2015})}\BibitemShut {NoStop}%
\bibitem [{\citenamefont {Tjhung}\ \emph {et~al.}(2018)\citenamefont {Tjhung},
  \citenamefont {Nardini},\ and\ \citenamefont {Cates}}]{Tjhung18}%
  \BibitemOpen
  \bibfield  {author} {\bibinfo {author} {\bibfnamefont {E.}~\bibnamefont
  {Tjhung}}, \bibinfo {author} {\bibfnamefont {C.}~\bibnamefont {Nardini}}, \
  and\ \bibinfo {author} {\bibfnamefont {M.~E.}\ \bibnamefont {Cates}},\
  }\href@noop {} {\bibfield  {journal} {\bibinfo  {journal} {Phys. Rev. X}\
  }\textbf {\bibinfo {volume} {8}},\ \bibinfo {pages} {031080} (\bibinfo {year}
  {2018})}\BibitemShut {NoStop}%
\bibitem [{\citenamefont {Nardini}\ \emph {et~al.}(2017)\citenamefont
  {Nardini}, \citenamefont {Fodor}, \citenamefont {Tjhung}, \citenamefont
  {Van~Wijland}, \citenamefont {Tailleur},\ and\ \citenamefont
  {Cates}}]{nardini2017entropy}%
  \BibitemOpen
  \bibfield  {author} {\bibinfo {author} {\bibfnamefont {C.}~\bibnamefont
  {Nardini}}, \bibinfo {author} {\bibfnamefont {{\'E}.}~\bibnamefont {Fodor}},
  \bibinfo {author} {\bibfnamefont {E.}~\bibnamefont {Tjhung}}, \bibinfo
  {author} {\bibfnamefont {F.}~\bibnamefont {Van~Wijland}}, \bibinfo {author}
  {\bibfnamefont {J.}~\bibnamefont {Tailleur}}, \ and\ \bibinfo {author}
  {\bibfnamefont {M.~E.}\ \bibnamefont {Cates}},\ }\href@noop {} {\bibfield
  {journal} {\bibinfo  {journal} {Phys. Rev. X}\ }\textbf {\bibinfo {volume}
  {7}},\ \bibinfo {pages} {021007} (\bibinfo {year} {2017})}\BibitemShut
  {NoStop}%
\bibitem [{\citenamefont {Caballero}\ and\ \citenamefont
  {Cates}(2020)}]{caballerocates}%
  \BibitemOpen
  \bibfield  {author} {\bibinfo {author} {\bibfnamefont {F.}~\bibnamefont
  {Caballero}}\ and\ \bibinfo {author} {\bibfnamefont {M.~E.}\ \bibnamefont
  {Cates}},\ }\href@noop {} {\bibfield  {journal} {\bibinfo  {journal} {Phys.
  Rev. Lett.}\ }\textbf {\bibinfo {volume} {124}},\ \bibinfo {pages} {240604}
  (\bibinfo {year} {2020})}\BibitemShut {NoStop}%
\bibitem [{\citenamefont {Paoluzzi}(2022)}]{paoluzzi2022scaling}%
  \BibitemOpen
  \bibfield  {author} {\bibinfo {author} {\bibfnamefont {M.}~\bibnamefont
  {Paoluzzi}},\ }\href@noop {} {\bibfield  {journal} {\bibinfo  {journal}
  {Phys. Rev. E}\ }\textbf {\bibinfo {volume} {105}},\ \bibinfo {pages}
  {044139} (\bibinfo {year} {2022})}\BibitemShut {NoStop}%
\bibitem [{\citenamefont {Theurkauff}\ \emph {et~al.}(2012)\citenamefont
  {Theurkauff}, \citenamefont {Cottin-Bizonne}, \citenamefont {Palacci},
  \citenamefont {Ybert},\ and\ \citenamefont {Bocquet}}]{Cecile2012}%
  \BibitemOpen
  \bibfield  {author} {\bibinfo {author} {\bibfnamefont {I.}~\bibnamefont
  {Theurkauff}}, \bibinfo {author} {\bibfnamefont {C.}~\bibnamefont
  {Cottin-Bizonne}}, \bibinfo {author} {\bibfnamefont {J.}~\bibnamefont
  {Palacci}}, \bibinfo {author} {\bibfnamefont {C.}~\bibnamefont {Ybert}}, \
  and\ \bibinfo {author} {\bibfnamefont {L.}~\bibnamefont {Bocquet}},\
  }\href@noop {} {\bibfield  {journal} {\bibinfo  {journal} {Phys. Rev. Lett.}\
  }\textbf {\bibinfo {volume} {108}},\ \bibinfo {pages} {268303} (\bibinfo
  {year} {2012})}\BibitemShut {NoStop}%
\bibitem [{\citenamefont {Buttinoni}\ \emph {et~al.}(2013)\citenamefont
  {Buttinoni}, \citenamefont {Bialk\'e}, \citenamefont {K\"ummel},
  \citenamefont {L\"owen}, \citenamefont {Bechinger},\ and\ \citenamefont
  {Speck}}]{Buttinoni13}%
  \BibitemOpen
  \bibfield  {author} {\bibinfo {author} {\bibfnamefont {I.}~\bibnamefont
  {Buttinoni}}, \bibinfo {author} {\bibfnamefont {J.}~\bibnamefont {Bialk\'e}},
  \bibinfo {author} {\bibfnamefont {F.}~\bibnamefont {K\"ummel}}, \bibinfo
  {author} {\bibfnamefont {H.}~\bibnamefont {L\"owen}}, \bibinfo {author}
  {\bibfnamefont {C.}~\bibnamefont {Bechinger}}, \ and\ \bibinfo {author}
  {\bibfnamefont {T.}~\bibnamefont {Speck}},\ }\href@noop {} {\bibfield
  {journal} {\bibinfo  {journal} {Phys. Rev. Lett.}\ }\textbf {\bibinfo
  {volume} {110}},\ \bibinfo {pages} {238301} (\bibinfo {year}
  {2013})}\BibitemShut {NoStop}%
\bibitem [{\citenamefont {Ginot}\ \emph {et~al.}(2018)\citenamefont {Ginot},
  \citenamefont {Theurkauff}, \citenamefont {Detcheverry}, \citenamefont
  {Ybert},\ and\ \citenamefont {Cottin-Bizonne}}]{Ginot2018}%
  \BibitemOpen
  \bibfield  {author} {\bibinfo {author} {\bibfnamefont {F.}~\bibnamefont
  {Ginot}}, \bibinfo {author} {\bibfnamefont {I.}~\bibnamefont {Theurkauff}},
  \bibinfo {author} {\bibnamefont {Detcheverry}}, \bibinfo {author}
  {\bibfnamefont {C.}~\bibnamefont {Ybert}}, \ and\ \bibinfo {author}
  {\bibfnamefont {C.}~\bibnamefont {Cottin-Bizonne}},\ }\href@noop {}
  {\bibfield  {journal} {\bibinfo  {journal} {Nat. Comm.}\ }\textbf {\bibinfo
  {volume} {9}},\ \bibinfo {pages} {696} (\bibinfo {year} {2018})}\BibitemShut
  {NoStop}%
\bibitem [{\citenamefont {van~der Linden}\ \emph {et~al.}(2019)\citenamefont
  {van~der Linden}, \citenamefont {Alexander}, \citenamefont {Aarts},\ and\
  \citenamefont {Dauchot}}]{vanderLinden19}%
  \BibitemOpen
  \bibfield  {author} {\bibinfo {author} {\bibfnamefont {M.~N.}\ \bibnamefont
  {van~der Linden}}, \bibinfo {author} {\bibfnamefont {L.~C.}\ \bibnamefont
  {Alexander}}, \bibinfo {author} {\bibfnamefont {D.~G. A.~L.}\ \bibnamefont
  {Aarts}}, \ and\ \bibinfo {author} {\bibfnamefont {O.}~\bibnamefont
  {Dauchot}},\ }\href@noop {} {\bibfield  {journal} {\bibinfo  {journal} {Phys.
  Rev. Lett.}\ }\textbf {\bibinfo {volume} {123}},\ \bibinfo {pages} {098001}
  (\bibinfo {year} {2019})}\BibitemShut {NoStop}%
\bibitem [{SM()}]{SM}%
  \BibitemOpen
  \href@noop {} {}\bibinfo {note} {See Supplementary Material at
  doi:}\BibitemShut {NoStop}%
\bibitem [{\citenamefont {Petrelli}\ \emph {et~al.}(2020)\citenamefont
  {Petrelli}, \citenamefont {Cugliandolo}, \citenamefont {Gonnella},\ and\
  \citenamefont {Suma}}]{Petrelli20}%
  \BibitemOpen
  \bibfield  {author} {\bibinfo {author} {\bibfnamefont {I.}~\bibnamefont
  {Petrelli}}, \bibinfo {author} {\bibfnamefont {L.~F.}\ \bibnamefont
  {Cugliandolo}}, \bibinfo {author} {\bibfnamefont {G.}~\bibnamefont
  {Gonnella}}, \ and\ \bibinfo {author} {\bibfnamefont {A.}~\bibnamefont
  {Suma}},\ }\href@noop {} {\bibfield  {journal} {\bibinfo  {journal} {Phys.
  Rev. E}\ }\textbf {\bibinfo {volume} {102}},\ \bibinfo {pages} {012609}
  (\bibinfo {year} {2020})}\BibitemShut {NoStop}%
\bibitem [{\citenamefont {Shi}\ \emph {et~al.}(2020)\citenamefont {Shi},
  \citenamefont {Fausti}, \citenamefont {Chat\'e}, \citenamefont {Nardini},\
  and\ \citenamefont {Solon}}]{Shi20}%
  \BibitemOpen
  \bibfield  {author} {\bibinfo {author} {\bibfnamefont {X.}~\bibnamefont
  {Shi}}, \bibinfo {author} {\bibfnamefont {G.}~\bibnamefont {Fausti}},
  \bibinfo {author} {\bibfnamefont {H.}~\bibnamefont {Chat\'e}}, \bibinfo
  {author} {\bibfnamefont {C.}~\bibnamefont {Nardini}}, \ and\ \bibinfo
  {author} {\bibfnamefont {A.}~\bibnamefont {Solon}},\ }\href@noop {}
  {\bibfield  {journal} {\bibinfo  {journal} {Phys. Rev. Lett.}\ }\textbf
  {\bibinfo {volume} {125}},\ \bibinfo {pages} {168001} (\bibinfo {year}
  {2020})}\BibitemShut {NoStop}%
\bibitem [{\citenamefont {Solon}\ and\ \citenamefont
  {Horowitz}(2022)}]{Solon22}%
  \BibitemOpen
  \bibfield  {author} {\bibinfo {author} {\bibfnamefont {A.}~\bibnamefont
  {Solon}}\ and\ \bibinfo {author} {\bibfnamefont {J.~M.}\ \bibnamefont
  {Horowitz}},\ }\href@noop {} {\bibfield  {journal} {\bibinfo  {journal} {J.
  Phys. A: Math. Theor.}\ }\textbf {\bibinfo {volume} {55}},\ \bibinfo {pages}
  {184002} (\bibinfo {year} {2022})}\BibitemShut {NoStop}%
\bibitem [{\citenamefont {Fullerton}\ and\ \citenamefont
  {Jack}(2016)}]{Jack16}%
  \BibitemOpen
  \bibfield  {author} {\bibinfo {author} {\bibfnamefont {C.~J.}\ \bibnamefont
  {Fullerton}}\ and\ \bibinfo {author} {\bibfnamefont {R.}~\bibnamefont
  {Jack}},\ }\href@noop {} {\bibfield  {journal} {\bibinfo  {journal} {J. Chem.
  Phys.}\ }\textbf {\bibinfo {volume} {145}},\ \bibinfo {pages} {244505}
  (\bibinfo {year} {2016})}\BibitemShut {NoStop}%
\bibitem [{\citenamefont {Sicilia}\ \emph {et~al.}(2007)\citenamefont
  {Sicilia}, \citenamefont {Arenzon}, \citenamefont {Bray},\ and\ \citenamefont
  {Cugliandolo}}]{Bray}%
  \BibitemOpen
  \bibfield  {author} {\bibinfo {author} {\bibfnamefont {A.}~\bibnamefont
  {Sicilia}}, \bibinfo {author} {\bibfnamefont {J.~J.}\ \bibnamefont
  {Arenzon}}, \bibinfo {author} {\bibfnamefont {A.~J.}\ \bibnamefont {Bray}}, \
  and\ \bibinfo {author} {\bibfnamefont {L.~F.}\ \bibnamefont {Cugliandolo}},\
  }\href@noop {} {\bibfield  {journal} {\bibinfo  {journal} {Phys. Rev. E}\
  }\textbf {\bibinfo {volume} {76}},\ \bibinfo {pages} {061116} (\bibinfo
  {year} {2007})}\BibitemShut {NoStop}%
\bibitem [{\citenamefont {Kolb}\ \emph {et~al.}(1983)\citenamefont {Kolb},
  \citenamefont {Botet},\ and\ \citenamefont {Jullien}}]{Kolb83}%
  \BibitemOpen
  \bibfield  {author} {\bibinfo {author} {\bibfnamefont {M.}~\bibnamefont
  {Kolb}}, \bibinfo {author} {\bibfnamefont {R.}~\bibnamefont {Botet}}, \ and\
  \bibinfo {author} {\bibfnamefont {R.}~\bibnamefont {Jullien}},\ }\href@noop
  {} {\bibfield  {journal} {\bibinfo  {journal} {Phys. Rev. Lett.}\ }\textbf
  {\bibinfo {volume} {51}},\ \bibinfo {pages} {1123} (\bibinfo {year}
  {1983})}\BibitemShut {NoStop}%
\bibitem [{\citenamefont {Meakin}(1983)}]{Meakin83}%
  \BibitemOpen
  \bibfield  {author} {\bibinfo {author} {\bibfnamefont {P.}~\bibnamefont
  {Meakin}},\ }\href@noop {} {\bibfield  {journal} {\bibinfo  {journal} {Phys.
  Rev. Lett.}\ }\textbf {\bibinfo {volume} {51}},\ \bibinfo {pages} {1119}
  (\bibinfo {year} {1983})}\BibitemShut {NoStop}%
\bibitem [{\citenamefont {Paul}\ \emph {et~al.}(2021)\citenamefont {Paul},
  \citenamefont {Bera},\ and\ \citenamefont {Das}}]{Paul21}%
  \BibitemOpen
  \bibfield  {author} {\bibinfo {author} {\bibfnamefont {S.}~\bibnamefont
  {Paul}}, \bibinfo {author} {\bibfnamefont {A.}~\bibnamefont {Bera}}, \ and\
  \bibinfo {author} {\bibfnamefont {S.~K.}\ \bibnamefont {Das}},\ }\href@noop
  {} {\bibfield  {journal} {\bibinfo  {journal} {Soft Matter}\ }\textbf
  {\bibinfo {volume} {17}},\ \bibinfo {pages} {645} (\bibinfo {year}
  {2021})}\BibitemShut {NoStop}%
\bibitem [{\citenamefont {Jullien}(1992)}]{Jullien92}%
  \BibitemOpen
  \bibfield  {author} {\bibinfo {author} {\bibfnamefont {R.}~\bibnamefont
  {Jullien}},\ }\href@noop {} {\bibfield  {journal} {\bibinfo  {journal}
  {Croatia Chemica Acta}\ }\textbf {\bibinfo {volume} {65}},\ \bibinfo {pages}
  {215} (\bibinfo {year} {1992})}\BibitemShut {NoStop}%
\bibitem [{\citenamefont {Thompson}\ \emph {et~al.}(2022)\citenamefont
  {Thompson}, \citenamefont {Aktulga}, \citenamefont {Berger}, \citenamefont
  {Bolintineanu}, \citenamefont {Brown}, \citenamefont {Crozier}, \citenamefont
  {in't Veld}, \citenamefont {Kohlmeyer}, \citenamefont {Moore}, \citenamefont
  {Nguyen} \emph {et~al.}}]{thompson2022lammps}%
  \BibitemOpen
  \bibfield  {author} {\bibinfo {author} {\bibfnamefont {A.~P.}\ \bibnamefont
  {Thompson}}, \bibinfo {author} {\bibfnamefont {H.~M.}\ \bibnamefont
  {Aktulga}}, \bibinfo {author} {\bibfnamefont {R.}~\bibnamefont {Berger}},
  \bibinfo {author} {\bibfnamefont {D.~S.}\ \bibnamefont {Bolintineanu}},
  \bibinfo {author} {\bibfnamefont {W.~M.}\ \bibnamefont {Brown}}, \bibinfo
  {author} {\bibfnamefont {P.~S.}\ \bibnamefont {Crozier}}, \bibinfo {author}
  {\bibfnamefont {P.~J.}\ \bibnamefont {in't Veld}}, \bibinfo {author}
  {\bibfnamefont {A.}~\bibnamefont {Kohlmeyer}}, \bibinfo {author}
  {\bibfnamefont {S.~G.}\ \bibnamefont {Moore}}, \bibinfo {author}
  {\bibfnamefont {T.~D.}\ \bibnamefont {Nguyen}},  \emph {et~al.},\ }\href@noop
  {} {\bibfield  {journal} {\bibinfo  {journal} {Computer Physics
  Communications}\ }\textbf {\bibinfo {volume} {271}},\ \bibinfo {pages}
  {108171} (\bibinfo {year} {2022})}\BibitemShut {NoStop}%
\bibitem [{\citenamefont {Ester}\ \emph {et~al.}(1996)\citenamefont {Ester},
  \citenamefont {Kriegel}, \citenamefont {Sander},\ and\ \citenamefont
  {Xu}}]{ester1996proceedings}%
  \BibitemOpen
  \bibfield  {author} {\bibinfo {author} {\bibfnamefont {M.}~\bibnamefont
  {Ester}}, \bibinfo {author} {\bibfnamefont {H.-P.}\ \bibnamefont {Kriegel}},
  \bibinfo {author} {\bibfnamefont {J.}~\bibnamefont {Sander}}, \ and\ \bibinfo
  {author} {\bibfnamefont {X.}~\bibnamefont {Xu}},\ }\href@noop {} {\
  (\bibinfo {year} {1996})}\BibitemShut {NoStop}%
\bibitem [{\citenamefont {Negro}\ \emph {et~al.}(2022)\citenamefont {Negro},
  \citenamefont {Caporusso}, \citenamefont {Digregorio}, \citenamefont
  {Gonnella}, \citenamefont {Lamura},\ and\ \citenamefont {Suma}}]{Negro2022}%
  \BibitemOpen
  \bibfield  {author} {\bibinfo {author} {\bibfnamefont {G.}~\bibnamefont
  {Negro}}, \bibinfo {author} {\bibfnamefont {C.~B.}\ \bibnamefont
  {Caporusso}}, \bibinfo {author} {\bibfnamefont {P.}~\bibnamefont
  {Digregorio}}, \bibinfo {author} {\bibfnamefont {G.}~\bibnamefont
  {Gonnella}}, \bibinfo {author} {\bibfnamefont {A.}~\bibnamefont {Lamura}}, \
  and\ \bibinfo {author} {\bibfnamefont {A.}~\bibnamefont {Suma}},\ }\href@noop
  {} {\bibfield  {journal} {\bibinfo  {journal} {Eur. Phys. J. E}\ }\textbf
  {\bibinfo {volume} {45}},\ \bibinfo {pages} {75} (\bibinfo {year}
  {2022})}\BibitemShut {NoStop}%
\bibitem [{mov()}]{movies}%
  \BibitemOpen
  \href@noop {} {}\bibinfo {howpublished} {Movies available at
  \url{https://www.dropbox.com/sh/lcx6wxwf91huhtp/AACHEq94Sb0isKKbsHNxs-Uka?dl=0}}\BibitemShut
  {NoStop}%
\bibitem [{\citenamefont {Alarc{\'o}n}\ \emph {et~al.}(2017)\citenamefont
  {Alarc{\'o}n}, \citenamefont {Valeriani},\ and\ \citenamefont
  {Pagonabarraga}}]{alarcon2017morphology}%
  \BibitemOpen
  \bibfield  {author} {\bibinfo {author} {\bibfnamefont {F.}~\bibnamefont
  {Alarc{\'o}n}}, \bibinfo {author} {\bibfnamefont {C.}~\bibnamefont
  {Valeriani}}, \ and\ \bibinfo {author} {\bibfnamefont {I.}~\bibnamefont
  {Pagonabarraga}},\ }\href@noop {} {\bibfield  {journal} {\bibinfo  {journal}
  {Soft matter}\ }\textbf {\bibinfo {volume} {13}},\ \bibinfo {pages} {814}
  (\bibinfo {year} {2017})}\BibitemShut {NoStop}%
\bibitem [{\citenamefont {Pohl}\ and\ \citenamefont
  {Stark}(2014)}]{pohl2014dynamic}%
  \BibitemOpen
  \bibfield  {author} {\bibinfo {author} {\bibfnamefont {O.}~\bibnamefont
  {Pohl}}\ and\ \bibinfo {author} {\bibfnamefont {H.}~\bibnamefont {Stark}},\
  }\href@noop {} {\bibfield  {journal} {\bibinfo  {journal} {Phys. Rev. Lett.}\
  }\textbf {\bibinfo {volume} {112}},\ \bibinfo {pages} {238303} (\bibinfo
  {year} {2014})}\BibitemShut {NoStop}%
\bibitem [{\citenamefont {Ma}\ and\ \citenamefont
  {Ni}(2022)}]{ma2022dynamical}%
  \BibitemOpen
  \bibfield  {author} {\bibinfo {author} {\bibfnamefont {Z.}~\bibnamefont
  {Ma}}\ and\ \bibinfo {author} {\bibfnamefont {R.}~\bibnamefont {Ni}},\
  }\href@noop {} {\bibfield  {journal} {\bibinfo  {journal} {The Journal of
  Chemical Physics}\ }\textbf {\bibinfo {volume} {156}},\ \bibinfo {pages}
  {021102} (\bibinfo {year} {2022})}\BibitemShut {NoStop}%
\bibitem [{\citenamefont {Cerd\`a}\ \emph {et~al.}(2004)\citenamefont
  {Cerd\`a}, \citenamefont {Sintes}, \citenamefont {Sorensen},\ and\
  \citenamefont {Chakrabarti}}]{cerda2004}%
  \BibitemOpen
  \bibfield  {author} {\bibinfo {author} {\bibfnamefont {J.~J.}\ \bibnamefont
  {Cerd\`a}}, \bibinfo {author} {\bibfnamefont {T.}~\bibnamefont {Sintes}},
  \bibinfo {author} {\bibfnamefont {C.~M.}\ \bibnamefont {Sorensen}}, \ and\
  \bibinfo {author} {\bibfnamefont {A.}~\bibnamefont {Chakrabarti}},\
  }\href@noop {} {\bibfield  {journal} {\bibinfo  {journal} {Phys. Rev. E}\
  }\textbf {\bibinfo {volume} {70}},\ \bibinfo {pages} {051405} (\bibinfo
  {year} {2004})}\BibitemShut {NoStop}%
\bibitem [{\citenamefont {Tartaglia}\ \emph {et~al.}(2018)\citenamefont
  {Tartaglia}, \citenamefont {Cugliandolo},\ and\ \citenamefont
  {Picco}}]{tartaglia2018coarsening}%
  \BibitemOpen
  \bibfield  {author} {\bibinfo {author} {\bibfnamefont {A.}~\bibnamefont
  {Tartaglia}}, \bibinfo {author} {\bibfnamefont {L.~F.}\ \bibnamefont
  {Cugliandolo}}, \ and\ \bibinfo {author} {\bibfnamefont {M.}~\bibnamefont
  {Picco}},\ }\href@noop {} {\bibfield  {journal} {\bibinfo  {journal} {J.
  Stat. Mech.}\ }\textbf {\bibinfo {volume} {2018}},\ \bibinfo {pages} {083202}
  (\bibinfo {year} {2018})}\BibitemShut {NoStop}%
\bibitem [{\citenamefont {Watanabe}\ \emph {et~al.}(2014)\citenamefont
  {Watanabe}, \citenamefont {Suzuki}, \citenamefont {Inaoka},\ and\
  \citenamefont {Ito}}]{watanabe2014ostwald}%
  \BibitemOpen
  \bibfield  {author} {\bibinfo {author} {\bibfnamefont {H.}~\bibnamefont
  {Watanabe}}, \bibinfo {author} {\bibfnamefont {M.}~\bibnamefont {Suzuki}},
  \bibinfo {author} {\bibfnamefont {H.}~\bibnamefont {Inaoka}}, \ and\ \bibinfo
  {author} {\bibfnamefont {N.}~\bibnamefont {Ito}},\ }\href@noop {} {\bibfield
  {journal} {\bibinfo  {journal} {J. Chem. Phys.}\ }\textbf {\bibinfo {volume}
  {141}},\ \bibinfo {pages} {234703} (\bibinfo {year} {2014})}\BibitemShut
  {NoStop}%
\end{thebibliography}%

\let\addcontentsline\oldaddcontentsline


\clearpage
\onecolumngrid

\begin{center}
    {\bf SUPPLEMENTAL MATERIAL}
\end{center}

\vspace{0.25cm}
\tableofcontents

\setcounter{equation}{0}
\renewcommand*{\theequation}{SM\arabic{equation}}

\setcounter{figure}{0}
\renewcommand*{\thefigure}{SM\arabic{figure}}

\section{A. Numerical Methods and Clusters Tracking Algorithm}

\subsection{A1. Numerical Methods}

We used a velocity Verlet algorithm in the $NVT$ ensemble that solves Newton’s equations of motion with the addition of two force terms,
a friction and a noise, which mimic a Langevin-type thermostat. Given the number of particles $N$ and the desired global packing fraction $\phi$, the 
surface $S$, fixed during the simulation, 
is obtained as $\phi= \pi \sigma^2 N/S$. For bulk simulations, particles are initialized with a random position. We use Molecular Dynamics (MD) time units, defined as $t_{\rm{MD}} = \sqrt{m\sigma_d^2/\epsilon}$, and fix the time-step of the simulations to  $d t = 0.001~ t_{\rm{MD}}$, which ensures numerical stability. 
To integrate the equation of motion, we used the open source software Large-scale Atomic/Molecular Massively 
Parallel Simulator (LAMMPS)~\cite{thompson2022lammps}, which enables to perform a simulation in parallel using multiple CPUs. On average, a typical simulation lasting $\approx 10^4~ t_{\rm{MD}}$ was run on 96 processors for a total of 72 hours on each CPU.  In the main text we do not write the time 
units explicitly.


\subsection{A2. Cluster Tracking Algorithm}
\label{sec:tracking}

We describe hereafter the numerical algorithm implemented to track the center
of mass of each cluster during the phase separating process, and follow their individual
motion, in the bulk or extracted from it, during the simulations.

At each time $t$, we use the DBSCAN clustering algorithm~\cite{ester1996proceedings} (also applied in~\cite{PRLino, Caporusso20, Negro2022})  to separate the particles
in $N_C(t)$ compact clusters, referred with labels $\mathcal{C}_\alpha$,
$\alpha=1, \dots, N_C$. The parameters for DBSCAN are the radius of
search around a particle, equal to $R=1.5\, \sigma$, $\sigma$ being very close to the particle's diameter
$d = 0.98\sigma$, and the minimum number of neighbours to assign particles to the same cluster $\epsilon=6$.
We checked that the clusters found by the algorithm are well enough preserved for small variations of these parameters.
Each cluster $\mathcal{C}_\alpha$ is composed by $N_\alpha = M_\alpha/m$ particles,
and in turn, each particle can be labelled with the cluster
$\mathcal{C}_\alpha$ it belongs to. This subdivision is then used
to compute all the per-cluster quantities presented in this work.

In order to study the evolution of each cluster over time, we need
an algorithm that identifies the same cluster at two successive times
$t$ and $t+\delta t$, with a fixed $\delta t = 10$ as time separation between two consecutive configurations of the system.
During such time window $\delta t$ the cluster may have lost or gained
some particles, it might have disaggregated or even fused with other clusters.
The tracking is achieved by considering the position of the cluster's center of mass at $t$,
and search at time $t+\delta t$ all clusters found by DBSCAN within
a radius $R=5\sigma$ from the initial center of mass position. Out of all these clusters,
we select
the one with the largest subset of particles in common with the initial one at time $t$, and with a number
of particles $N_{\beta}(t+\delta t)$ in between 0.9 and 1.1 times the original number of particles $N_{\alpha}(t)$.
If there is no cluster
satisfying these two conditions, the tracking of the selected cluster stops.
Consequently, the number of tracked clusters diminishes with time.

\begin{figure}[h!]
    \includegraphics[scale=1.1]{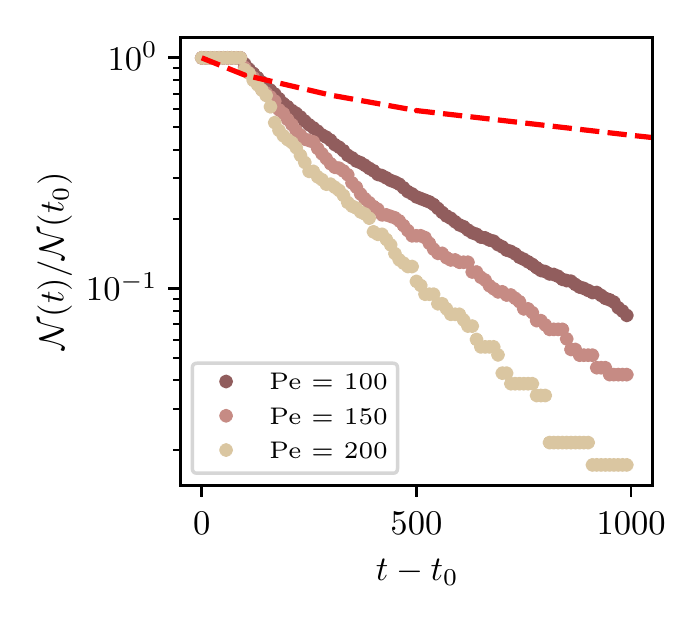}
    \vspace{-0.35cm}
    \caption{
        {\bf Evolution of the total number
                of clusters}, ${\cal N }(t)$, followed by the tracking algorithm at time $t$, as described in the text,
        normalized by the total number of clusters identified at the initial time of the tracking, ${\mathcal N} (t_0)$.
        The dashed red line is $N_C(t)/N_C(t_0)$ for Pe = 100, where $N_C(t)$ is the instantaneous
            number of clusters, independently of their history.
    }
    \label{fig:tracking_ncluster_time}
\end{figure}

As explained above, we lose clusters along the evolution, and the number of successfully tracked clusters decreases in time.
We report in Fig.~\ref{fig:tracking_ncluster_time} the number of clusters
${\cal N}(t)$ still tracked by the algorithm during the evolution of large systems,
when clusters coalesce and grow, divided by the number of clusters found at the initial time $t_0$,
${\cal N}(t_0)=N_C(t_0)$. This fraction decreases with an approximately exponential law, and its characteristic
time gives an estimation of the typical time required for the clusters to collide or disgregate. We note that ${\cal N}(t)$
is slightly different from the instantaneous number of clusters identified by DBSCAN independently of their
history, $N_C(t)$, which is plotted with a red dashed line in Fig.~\ref{fig:tracking_ncluster_time} 
and is also reported in the main text in Fig.~5(b).

After performing the tracking of each cluster over time, we are able
to follow the trajectories of their centers of mass, ${\bf r}^\alpha_{\rm cm}(t)$, and from them
compute the mean square displacement and other observables. The individual trajectories
${\bf r}^\alpha_{\rm cm}(t)$, and hence the displacements
$({\bf r}^\alpha_{\rm cm}(t)-{\bf r}^\alpha_{\rm cm}(t_0))^2$, last until the tracking is able to
follow the selected cluster. The mean-square displacement is computed as an average over all these
trajectories and at each time $t$ the normalization is given by the corresponding ${\cal N}(t)$, which
decreases with $t$.

In order to study  in more depth the mechanism with which the clusters grow, and get a grip on the collision
events between clusters, we also use a modification of the previous algorithm. 
We compare two configurations at two successive times $t$ and $t+\delta t$, and we establish i) a correspondence between clusters, and ii) the type of event the clusters have been involved in, based on the computation of the overlap between them.
More precisely, the normalized overlap between cluster $i$ at time $t$ and cluster $j$ at time $t+\delta t$ is defined as
\begin{equation}
  o_{ij}=\frac{n_{ij}}{\sqrt{M_i M_j}} \mbox{,}
\end{equation}
where $n_{ij}$ is the number of particles belonging to both clusters and $M_i$ and $M_j$ are the masses of the two clusters.
The overlap is computed for all $i=1,...,\bar{N}_C(t)$ and $j=1,...,\bar{N}_C(t+\delta t)$, with $\bar{N}_C$ the number of clusters of mass $M>100$.
We have identified this mass threshold as the one that separates stable long-living clusters from instantaneous aggregations of particles.

As we find a non zero overlap $o_{ij}>0$, we check the mass variation $\Delta M_{ij}$ between the cluster $i$ at time $t$ and the cluster $j$ at time $t+\delta t$.
\begin{itemize}
\item If $|\Delta M_{ij}|<100$, then only evaporation or condensation of particles have occurred.
The cluster labelled $i$ at time $t$ has not undergone any collisions, and it is the same cluster, labelled $j$, at time $t+\delta t$.
All these $N_p$ events are recorded as events due to Ostwald ripening.
In the case one of this events is recorded, any other overlap involving cluster $i$ or $j$ is disregarded.
\item All the other $N_{coll}$ occurrences of $o_{ij}>0$ and $|\Delta M_{ij}| >100$ are recorded as 
collisions between clusters or break ups of clusters in pieces. 
\end{itemize}

\section{B. Movies}

We complement the information given in the figures with a series of movies \cite{movies}.

\begin{itemize}
    \item Movie M1.mp4 displays the motion of an active cluster extracted from the bulk (as explained in the main text)
          on a large timescale, which highlights the cluster's movement, and
          on a short timescale, in which the local correlations of the force are shown.
    \item In movie~M2.mp4   we show the evolution of the active system with Pe = 100. We
          painted the regular clusters with fractal dimension $d_f=2$ in blue,  and the fractal clusters with
          dimension $d_f<2$ in gray. The film demonstrates the coexistence of compact and fractal clusters at all times. Several
          interesting features are also exhibited in the movie:
          \begin{itemize}
              \item
                    Some small clusters progressively evaporate and disappear, as in Ostwald ripening.
              \item
                    The aggregation of already large clusters leads to the formation of even larger and fractal structures,
                    as in diffusion-aggregation processes.
              \item
                    The clusters have irregular interfaces.
              \item
                    At late times  a few very large elongated clusters have been formed and only a few small regular ones appear and
                    disappear.
          \end{itemize}
    \item
          In movie~M3.mp4 we show the evolution of the passive system 
          (see Sec.~E of the SM for further information about the parameters used) and we
          painted the clusters as in movie~M2: regular clusters with fractal dimension $d_f=2$ in blue,  and fractal clusters with
          dimension $d_f<2$ in gray. Again, there is coexistence of compact and fractal clusters at all times. However, the
          configurations are very different from the ones in movie~M2:
          \begin{itemize}
              \item
                    The clusters are generically rounder and globally smaller than in the active case.
              \item
                    Until the end of the evolution there are many regular clusters with $d_f=2$.
          \end{itemize}
    \item
          In movie M4.mp4 we show, side-by-side, the evolution of two clusters extracted from the bulk:
          an active one on the left and a passive one on the right. Note that the time scales in the two videos (shown in the upper
          right keys) are not the same, the one
          for the passive cluster is much longer. The c.o.m. displacement is drawn in red in both panels.
          The color code represents the modulus of the local hexatic order.
          The differences are apparent. The active cluster is much more mobile and changes form while the passive
          one remains globally quite undeformed over a much longer period, with just some rearrangements at its surface.
          The (rather straight) light colored regions within the active cluster are located on the interfaces between patches of different
          orientational order, they do not mean that the cluster broke, only that there are ever-changing smaller structures within
          it.

\end{itemize}

\section{C. Further numerical analysis of the active system}

In this Section we present complementary numerical results.

\subsection{C1. Scaling of the averaged cluster mass}

\begin{center}
    \begin{figure}[h!]
        \includegraphics[scale=1]{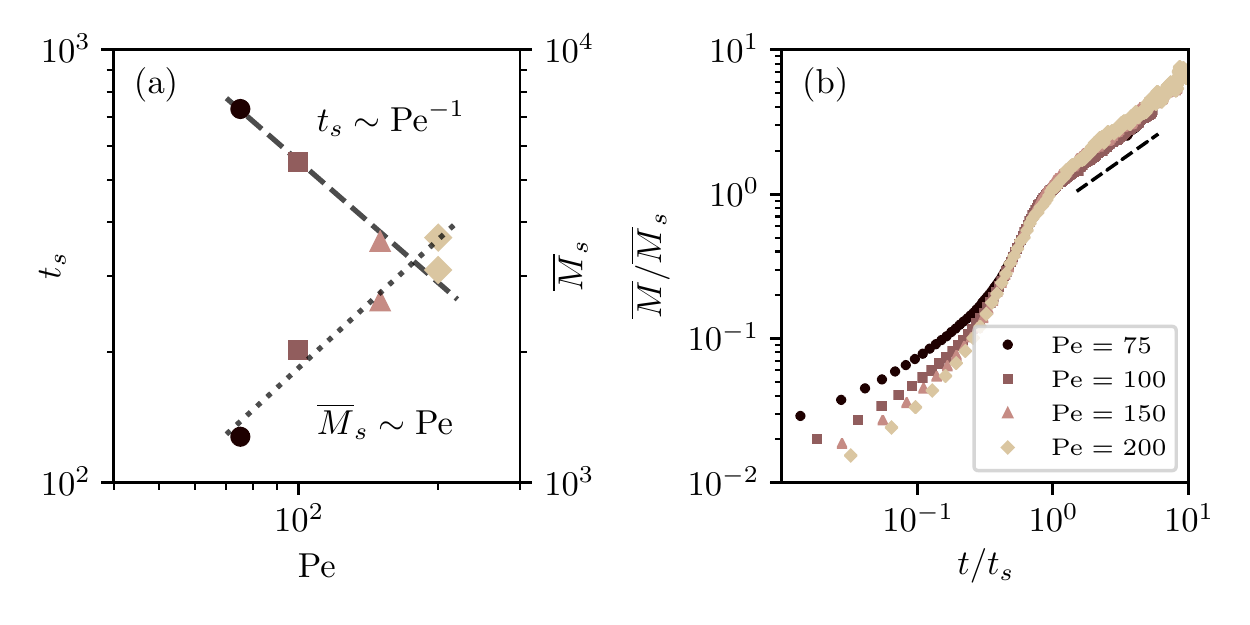}
        \vspace{-0.5cm}
        \caption{{\bf Scaling of the averaged cluster mass.}
            (a) The time needed to enter the scaling regime, $t_s$, against Pe, with a fit of the power law decay Pe$^{-1}$, and
            the mass at the scaling time $t_s$ with the power law growth proportional to Pe. See the text for the details.
            (b) Scaling of the averaged mass,
            using $t_s$ and $\overline M_s$: $\overline M/\overline M_s = f(t/t_s)$ with $f(1)=1$ and $f(x) \sim
                x^{2/3}$ for $x>1$ and before saturation (dashed segment close to the data). Due to the Pe dependence of
            $t_s$ and $\overline M_s$, the time dependent mass follows $\overline M = \mbox{Pe}  \,  \overline f(t \, \mbox{Pe})$
            with $\overline f(1)$ finite and $\overline f(t \, \mbox{Pe}) \sim (t \, \mbox{Pe})^{2/3}$ for $t> t_s$ which implies
            $\overline M(t>t_s) \sim \mbox{Pe}^{5/3}$ at fixed $t$.
        }
        \label{fig:scaling}
    \end{figure}
\end{center}

Figure~1 of the main text shows the time-dependence of the averaged
cluster mass, $\overline M$, for several values of Pe.
We estimate the time $t_s$ at which the system enters the scaling regime from the
horizontal position of the shoulder, after which the asymptotic $t^{2/3}$ behavior establishes.
In Fig.~\ref{fig:scaling}(a) we plot the Pe dependence of $t_s$ extracted in this way.
An algebraic fit to the numerical data yields
Pe$^{-0.93}$. We therefore assume that the decay is
$t_s \sim ({\rm Pe}-a)^{-1}$,
as this time should diverge at a finite Pe $>0$, the limit of the MIPS sector of the
phase diagram. The fit of the exponent is shown in Fig.~\ref{fig:scaling}(a).
Concomitantly, the mass at this time, $\overline M_s$, also shown in this plot,
is fitted by a growing function of Pe as $ ({\rm Pe}-b)^{0.98}$, and
we will assume it simply goes as $\overline M_s \sim {\rm Pe}$.

In Fig.~\ref{fig:scaling}(b) we perform a scaling of the $\overline M$ data using $t_s$, as previously determined,
and the value of $\overline M_s$ extracted from the
vertical position of the shoulder in Fig.~1 of the main text, $\overline M_s= \overline M(t_s)$,
see Fig.~\ref{fig:scaling}(a).
The scaling is quite satisfactory both in the  long-time (scaling) regime in which $\overline M$ grows as
$t^{2/3}$ (black dashed line in the figure)
as well as in the previous regime in which the growth is faster. The scaling implies that the dependence
of the time-dependent averaged mass with
Pe is $\overline M = \mbox{Pe}  \,  \overline f(t \, \mbox{Pe})$
with $\overline f(1)$ finite and $\overline f(t \, \mbox{Pe}) \sim (t \, \mbox{Pe})^{2/3}$ for $t> t_s$ which implies
$\overline M(t>t_s) \sim \mbox{Pe}^{5/3}$ at fixed $t$. We checked that this Pe dependence complies with the data
at long times, with a fit yielding an exponent of 1.67 (not shown).

An averaged cluster mass $\overline M$ growing linearly with Pe was found in the steady state of the Janus particle suspension at intermediate
densities studied in~\cite{Cecile2012}. Simulations of various active particle models which also reach steady states
often showed the opposite trend~\cite{alarcon2017morphology,pohl2014dynamic}.

\subsection{C2. Mass and radius of gyration distributions}

As suggested by the two inserts in Fig.~1 of the main paper, clusters of different sizes and shapes are present during the coarsening
process towards full phase separation.
In particular, in the time regime in which dynamical scaling is verified, $t>t_s$, we observe small compact clusters of rounded shapes
and large fractal clusters coexisting  (see also Fig.~5 of the main text).
The evolution in time of the (un-normalised) cluster mass distribution in a system with Pe = 100 is reported in the right inset in Fig.~1 of the main text, where the plot x-axis is limited to only relatively small masses (similar results can be obtained at other values of Pe.)
Here we analyze the mass and gyration radius distributions on much larger scales, showing that
the form of these distributions is compatible with the presence of two separate length scales
linked to the two types of clusters with different fractal dimensions.


\begin{figure}[h!]
    \centering
    \includegraphics{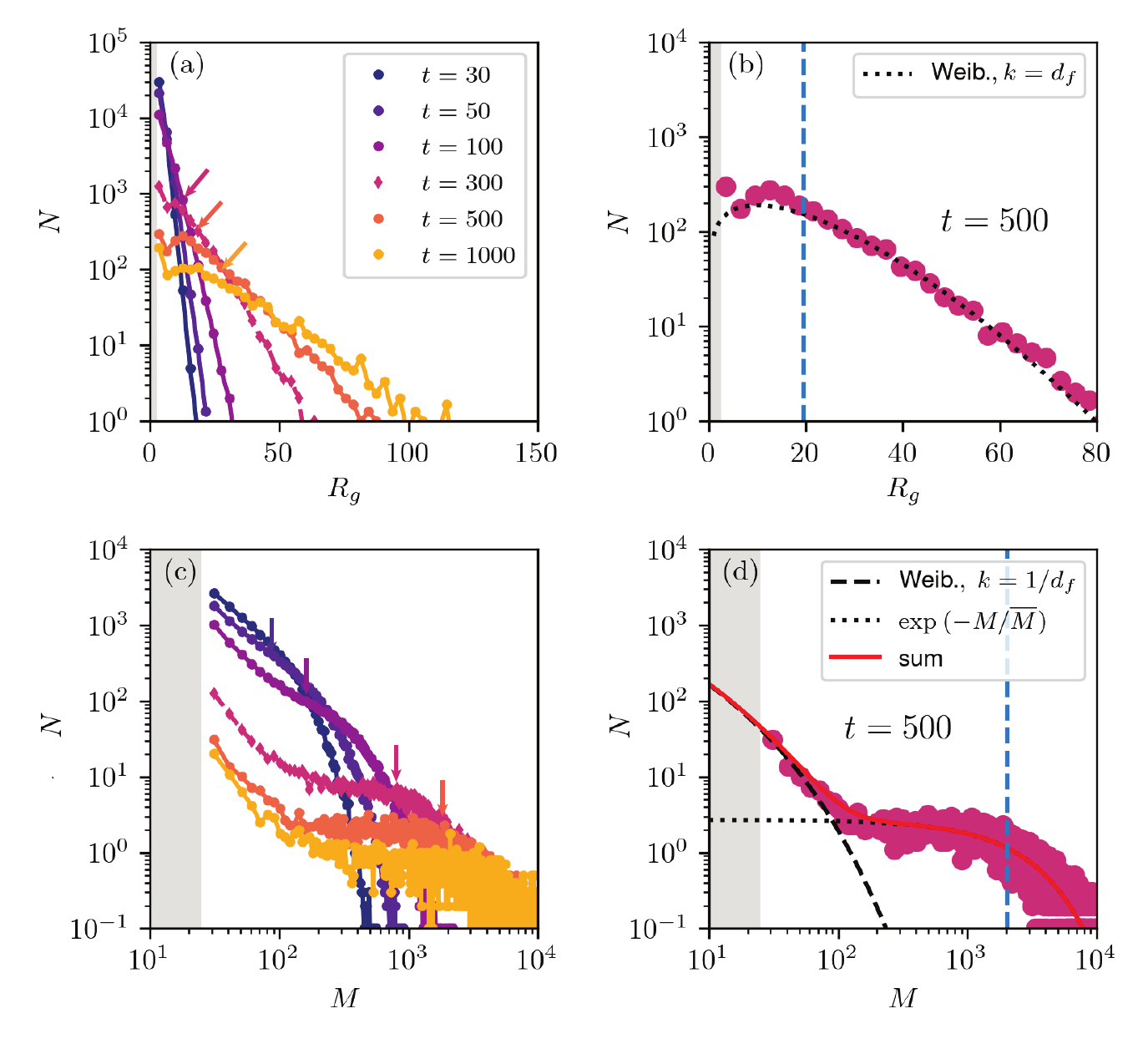}
    \caption{{\bf Radius of gyration and mass distributions in an active system} at Pe $=100$ and $\phi=0.5$. $t_s \sim 300$ is approximately the beginning of the scaling regime for these parameters.
    (a) Histogram of the radius of gyration for all clusters at several times within the scaling regime.
    The arrows indicate the global averages. Here and in the other panels the gray regions are excluded by the cut-off $M>25$ that we imposed on the size of the clusters taken into account.
    (b) The data at time $t=500$ together with a fit to a Weibull form $N(R_g) = R_g^{k-1} e^{-(R_g/R_g^*)^k}$ with $k=1.45$ which we
    will interpret as the fractal dimension of large clusters, and
    $R_g^* \sim 16$, which is very close to the average $\overline R_g$ at this time
    shown with a vertical dashed (blue).
    (c) Histogram of the cluster masses at the same times as in (a).
    The arrows indicate the global averages.
    (d) Fit of the histogram of clusters masses at $t=500$ with a Weibull distribution with $k=1/d^s_f=1/2$, see Eq.~(\ref{eq:two_weibull})
    dashed line describing the regime $M< \overline M$, added to a pure exponential (dotted line) describing the regime $M>\overline M$.
    The red curve shows the sum of the two terms, with parameters $c_s = 2377$,  $M_s^* = 4.3$, $c_l =  2.7$,  and $M_l^* = 2374$
    which is very close to $\overline M = 2019$.
    These plots should be compared to the ones in
    Fig.~\ref{fig:dist-Rg-attractive} where the same distributions for the passive model are studied.
    }
    \label{fig:CSDs}
\end{figure}

First of all, in Fig.~\ref{fig:CSDs}(a) and (b) we show the histograms $N(R_g)$. The radius of gyration of the $\alpha$ cluster is computed as
\begin{equation}
    R^\alpha_g = \sqrt{\frac{m}{M_\alpha} \sum_{i=1}^{M_\alpha/m} (\mathbf{r}^\alpha_i-\mathbf{r}^\alpha_{\rm cm})^2}
    \; .
\end{equation}
$M_\alpha/m$ is the number of particles in the $\alpha$th cluster and $\mathbf{r}^\alpha_{\rm cm}$
is the center of mass of the same cluster. In (a) we see how the histograms get wider at longer times.
After entering the scaling regime at $t_s\sim 300$, the data flatten, and even develop a shallow minimum,
for small $R_g$. This suggests that the
seeming plateau in $N(M)$ in Fig.~1 of the main text is just the beginning of a second slower regime, as confirmed in
panels (c) and (d). The global average $\overline{R}_g$
at each time is indicated with inclined arrows in (a), with the color of the corresponding full curve,  and with a vertical dashed blue line in (b).
It falls roughly at the beginning of the  second regime.

We propose that the distribution of the radii of gyration in the tail of the curve (for large $R_g$) takes the Weibull form
\begin{equation}
    \label{eq:rg_exp_distribution}
    N(R_g)\sim R^{k-1}_g e^{-(R_g/R_g^*)^k}
    \; \mbox{.}
\end{equation}
with parameter $k$. Figure \ref{fig:CSDs}(b) shows $N(R_g)$ at $t=500$ together with this fit (done for $R_g>30$).
The value of  $R_g^*$ extracted from the fit is very close to the averaged value $\overline R_g$. The value of $k$ obtained is $1.45$, which is equal to the fractal dimension obtained for large masses $d^l_f=1.45$ (see Fig.~5 of the main text). We can thus identify here $k=d^l_f$.
Note that the data at small $R_g$ are not well fitted by this curve, as they have a different fractal dimension $d^s_f=2$, and potentially follow a different law. At the same time, $N(R_g)$ is not well suited to see which law is followed for small masses, as the latter are represented by few points. 
Instead, we show below that the $N(M)$ distribution is better suited for the purpose and allows us to better identify the complete functional form which includes both small and large masses. 

In order to obtain a functional form of the mass distribution $N(M)$ that works for both small and large masses, we can start from the functional form of the radii distribution expected for each case, 
and transform them separately in the relative mass distribution using the relation $M \sim R_g^{d_f}$, with the fractal dimension being either $d^s_f$ or $d^l_f$ depending on the case.
For large masses, we can use Eq.~(\ref{eq:rg_exp_distribution}), which correspond to an exponential mass distribution $e^{-M/M^*_l}$. 
For small masses, we find that the best function is an exponential $e^{-R_g/R_{gs}^*}$, which correspond to a Weibull distribution for the masses. 
Thus, the total mass distribution is given by a superposition of the two following forms:
\begin{eqnarray}
    \label{eq:two_weibull}
    N(M) &\sim&
    c_s \, M^{ (1/d_f^s-1)} \, e^{ -(M/M^*_s)^{1/d_f^s} }
    + c_l  \, e^{-(M/M^*_l)}
    \; .
\end{eqnarray}
The two constants $c_s$ and $c_l$ and the two masses $M^*_l$ and $M^*_s$ are left as free parameters of the fit. As shown in Fig.~\ref{fig:CSDs}(d), the distribution form is fully consistent with the histogram of the cluster masses. The values of $M^*_s$  and $M^*_l$  returned from the fit are reasonable and the latter is compatible with the full average $\overline M$.
Note that this procedure is strictly valid only if the distribution is defined piecewise over two disjoint intervals of $R_g$,
$[0:R_<]$ and $[R_>:\infty]$, with $R_< < R_>$.
Given the large separation between the average masses for the two types of clusters and the consistency of the fit, we believe that applying the transformations separately
(and not writing theta functions explicitly) is a good approximation.

The abundance of the two species of clusters, fractal and compact, over the time evolution of the system is reported in Table~\ref{tab:mass_proportion}.

\begin{table}[h!]
    \begin{tabular}{ |c|c|c| }
        \hline
        \, time \, & \, \% $(M>\overline M)$ \, & \, \% $(M<\overline M)$ \,
        \\
        \hline
        \hline
        10         & 0.375                      & 0.625
        \\
        \hline
        30         & 0.350                      & 0.650
        \\
        \hline
        50         & 0.343                      & 0.657
        \\
        \hline
        100        & 0.337                      & 0.662
        \\
        \hline
        300        & 0.378                      & 0.622
        \\
        \hline
        500        & 0.371                      & 0.629
        \\
        \hline
        600        & 0.361                      & 0.639
        \\
        \hline
        800        & 0.350                      & 0.650
        \\
        \hline
        1000       & 0.353                      & 0.648
        \\
        \hline
        3000       & 0.340                      & 0.660
        \\
        \hline
    \end{tabular}
    \caption{Details on the cluster mass (and radius of gyration) histograms.
        We show the fraction of clusters belonging to the two families identified according to the value of their fractal dimension.
    }
    \label{tab:mass_proportion}
\end{table}

Finite-size clusters have been observed in experiments of active Janus colloids in~\cite{Ginot2018}, in a segregated steady state where the dynamics proceeds through an aggregation-fragmentation process involving only single particles. Clusters are small compact objects of regular shape, while the formation of large aggregates is suppressed in the regime of activity considered in this experiment. Therefore, a second length scale is not present
and a single exponential distribution was measured.
Other systems of non scalar active particles where MIPS is arrested, are characterized by a different cluster size distribution. For instance, in~\cite{vanderLinden19} and~\cite{ma2022dynamical}
systems of aligning and chiral active particles were analyzed and a truncated power law was observed for the cluster size distribution in the stationary regime. From a comparison with our results, we conclude that both alignment and intrinsic torque (present in these systems but not in ours)
promote a broad spectrum of lengths for the cluster sizes, rather than a single well defined one.

\subsection{C3. Clusters' diffusion coefficient}

We describe in this Section the method that we employed to estimate the diffusion
coefficient of the clusters, from the trajectories of their centers of mass.

\begin{figure}[h!]
    \includegraphics[scale=1.2]{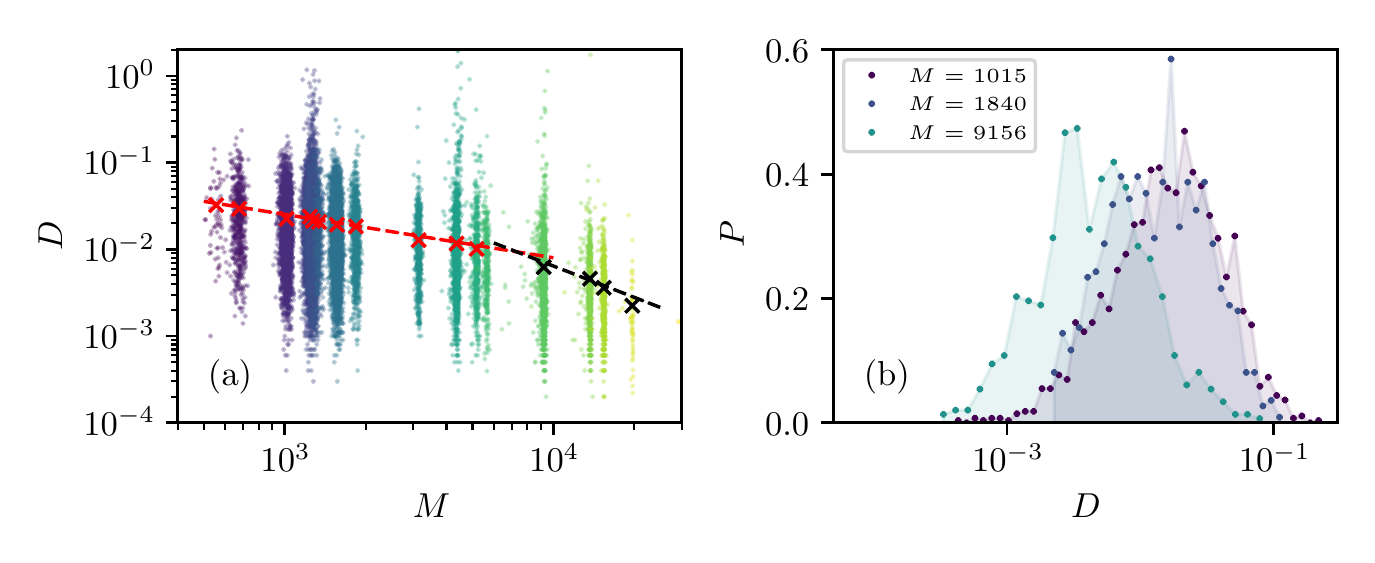}
    \vspace{-0.5cm}
    \caption{
        {\bf Clusters' diffusion coefficient.} Analysis of the  diffusion coefficient of the centers of mass  of the clusters
        extracted from the bulk. (a) Scatter plot of the individual diffusion coefficient and the associated average mass over the trajectory.
        (b) Distribution of the diffusion coefficients of a cluster started with the same initial condition and
        evolved independently in different runs that give rise to the histograms. The parameters are the same as in other figures, Pe = 100 and $\phi=0.5$.
    }
    \label{fig:diff-coeff}
\end{figure}

Compared to single particles, clusters undergo larger mass fluctuations over their evolution, which can result in
jumps in the clusters displacements. For this reason, as already mentioned in the
SM Sec.~1,
we follow clusters only if their mass fluctuation after the time evolution $t+\delta t$ is within $10\%$ of the mass at the time $t$.

After obtaining a set of trajectories ${\mathbf r}_{\text{cm}}^{\alpha}$ for the center of mass of each cluster $\alpha$,
with average mass $M_\alpha$ (over the trajectory), we extract an estimate of the diffusion coefficient via the formula:
\begin{equation}
    D_{\alpha} = \frac{1}{2Nd}\sum_{k=0}^{N} \frac{({\mathbf r}_{\text{cm}}^{\alpha}(t_{k})-{\mathbf r}_{\text{cm}}^{\alpha}(t_{0}))^2}{(t_{k}-t_0)}
    \; ,
    \label{eq:diff_coeff_discrete}
\end{equation}
where $t_i$ are the discretized times, with the initial time of the cluster's trajectory being $t = t_0$, and the sum extended up to the last time $t = t_N$.
In this way we obtain a set of $D_\alpha$ with associated mass $M_\alpha$.
We can then bin appropriately the range of masses, and for each bin compute the average value of the diffusion coefficient,
obtaining  the diffusion coefficient of the clusters $D$ as a function of the mass $M$.
For clusters tracked in the whole system (Fig.~\ref{fig:dos} of the main text),
the average values of $D$ are computed by a logarithmic binning of the respective masses, while for isolated clusters,
we average over several independent realizations of the same initial condition.

Figure \ref{fig:diff-coeff}, left panel, shows the scatter plot (circles) of $D_\alpha$ as a function of $M_\alpha$ for extracted clusters (see main text).
The crosses represent the average values of $D$ obtained after binning. The dashed lines correspond to fits of $D(M)$ for $M<5000$,
via $D(M) \sim M^{-1/2}$ (red curve), and for $M>5000$ via $D(M) \sim M^{-1}$ (black).
Figure \ref{fig:diff-coeff}, right panel, shows the probability distribution  of $D$ for some averaged masses
given in the key.

\subsection{C4. Cluster Displacements}

In this section, we show the probability distribution function of
the displacements $\delta r^{\text{cm}}_a = r^{\text{cm}}_a(t) - r^{\text{cm}}_a(t_0)$, with $a = 1,2$, respectively, the $x$ and $y$ Cartesian components.
The displacements are computed by numerical simulation for the c.o.m. trajectory of isolated clusters in the time interval $\delta t = t - t_0$,
where, as usual, $t_0$ is the initial time of the tracking.
In order to improve the statistics, we gather together the two displacement components, 
and we compute the cumulative probability distribution function. In the figure, we denote simply as $\delta r_a^{\rm cm}$ these gathered Cartesian displacements.

\begin{figure}[h!]
    \centering
    \includegraphics[scale=0.8]{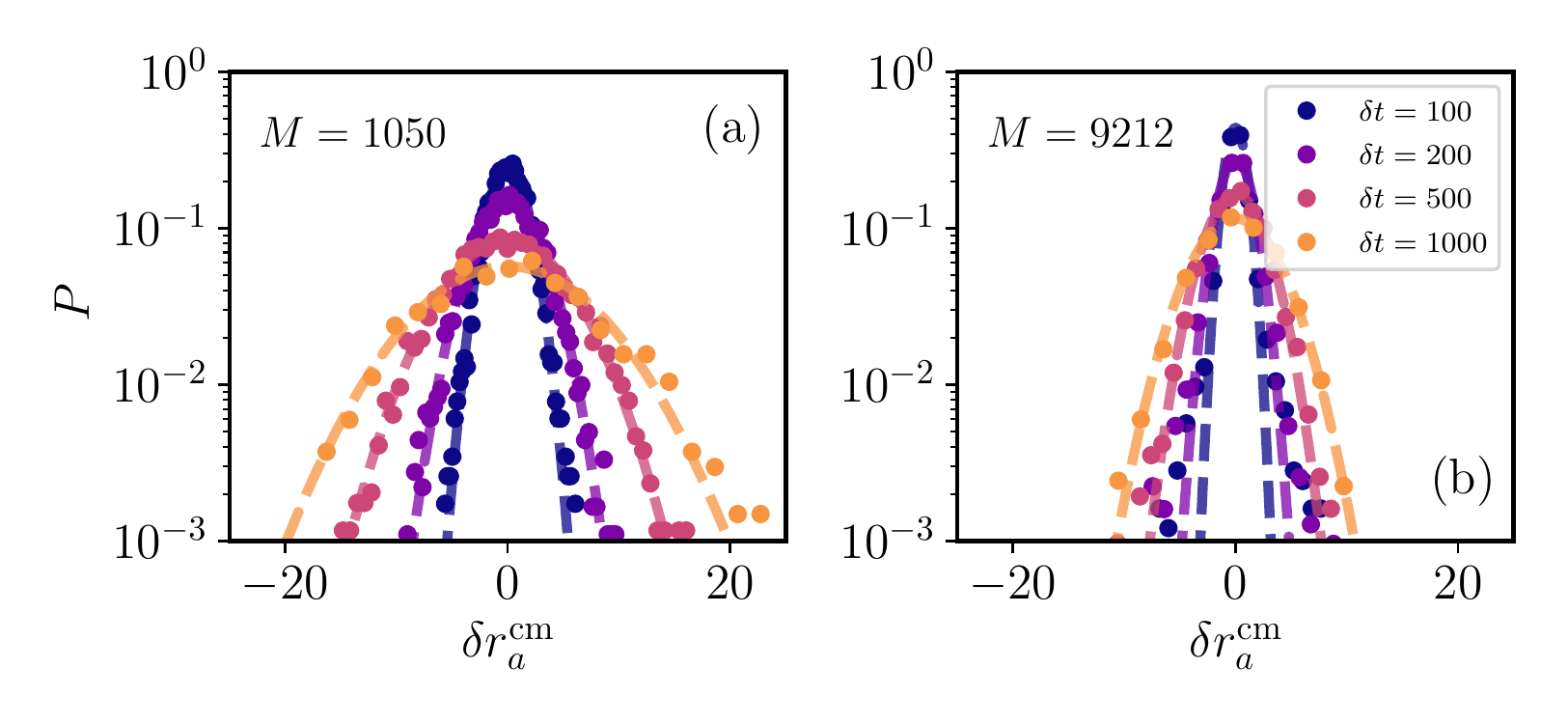}
    \vspace{-0.5cm}
    \caption{{\bf Distribution of displacements} $\delta r^{\rm cm}_a$ along the two Cartesian 
    directions at different tracking times $\delta t$ (in the key).
        In one case, the cluster is such that
        $D \sim M^{-0.5}$ regime (left), and in the other it corresponds to the $D \sim M^{-1.0}$ regime (right).
        The stationary average masses over all the different evolutions of the same initial condition are reported in top left part of
        the two panels.
    }
    \label{fig:cluster_displacement}
\end{figure}

The results for two initial masses and different time delays are shown in Fig.~\ref{fig:cluster_displacement}. Panel (a) displays
data for clusters with an averaged mass  $M \approx 1050$
that is clusters such
that the diffusion coefficient goes as $D \sim M^{-0.5}$.
(On average, the cluster's mass increases a little bit
until reaching stationary conditions at a new value in the extraction procedure.)
Instead, panel (b) presents data
for larger clusters, representative of the
$D \sim M^{-1}$ regime. All data are satisfactorily fitted
with Gaussian distributions, and we did not find any significant difference between the small and large mass cases.
(Some outliers  for very large masses may not be sufficiently well sampled.)

These measurements confirm that smaller clusters move, in general, more than larger ones, as demonstrated by the broader
distribution of displacements in panel (a) than in panel (b), at equal tracking times.

\subsection{C5. Forces on individual clusters}

Each cluster is subject to a net potential force, a net active force, and a total torque.

The total potential force vanishes identically due to the action-reaction principle. 

\begin{figure}[h!]
    \vspace{0.25cm}
    \centering
    \includegraphics[width=17cm]{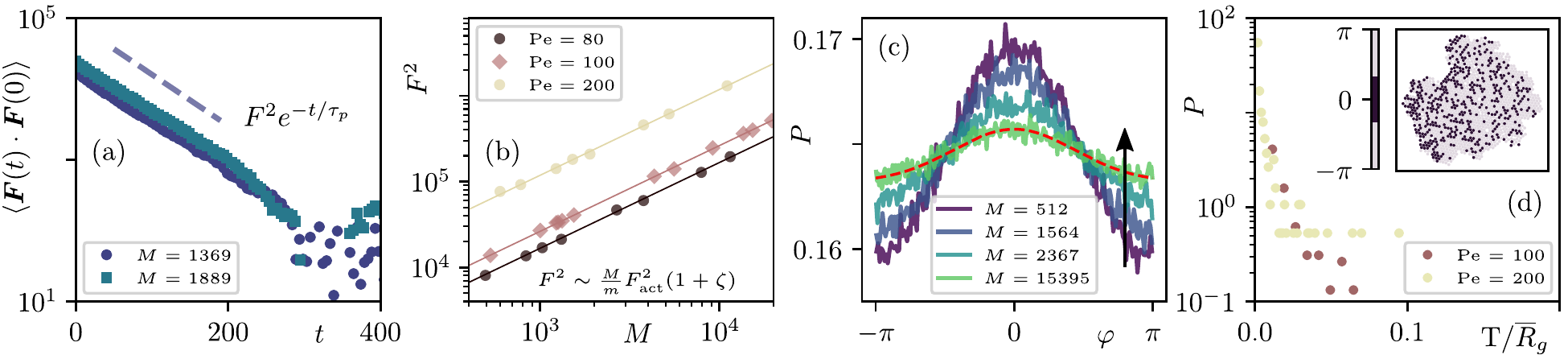}
    \caption{
        {\bf Forces on individual clusters extracted from the bulk.}
        (a) Exponential decay, with
        characteristic time $\tau_p \approx 60$,  of the correlation of the total active force acting on the clusters, Pe = 100.
        (b) Equal time value as a function of the cluster mass $M$.
        These curves confirm the $F_{\rm act}^2$ dependence, a fit yields
        $\zeta=0.05$,
        and extrapolate to $F_{\rm act}^2$ for $M\to m$.
        (c) Distribution of the angles between the ${\bf n}_i$
        and $\bm{F}= F_{\rm act} \sum_{j=1}^{M/m} {\bf n}_j$.
        (d)  Distribution of the total torque ${\rm T}$ normalized by the averaged radius of gyration.
        Inset: map of the local active force alignment with the direction of motion.}
    \label{fig:quattro}
\end{figure}

The total active force, instead, does not. It is the vectorial sum of the individual active forces
and equals $\bm{F}= F_{\rm act} \sum_{i=1}^{M/m}  {\bf n}_i$.
Its temporal correlations are shown in Fig.~\ref{fig:quattro}(a) for clusters extracted from the bulk, and 
for all masses studied they decay exponentially, with a characteristic time $\approx\tau_p$,
that is of the order of the persistence time of the single active particle.
At equal times
$F^2 \equiv \langle \bm{F}^2(0)\rangle =  F^2_{\rm act} \, [ \sum_{i=1}^{M/m}  \langle {\bf n}^2_i(0) \rangle + \sum_{i\neq j}^{M/m}  \langle {\bf n}_i(0) \cdot {\bf n}_j(0)\rangle ]$ and, 
assuming that the ${\bf n}_i(0)$ have independently identically distributed ({\it i.i.d.}) 
random components, $F^2 \equiv \langle \bm{F}^2(0)\rangle \sim (M/m) F_{\rm act}^2 ( 1 + \zeta)$. 
The measurement on clusters extracted from the bulk is in good agreement with this assumption, with
$\zeta \approx 0.05$, see Fig.~\ref{fig:quattro}(b).

However, the active forces are correlated and the components of the ${\bf n}_i(0)$ are not 
{\it i.i.d.}. This is proven by the pdf of the angle $\varphi$ between   ${\bf n}_i$
and  $\bm{F}$, which is shown in Fig.~\ref{fig:quattro}(c) as measured  in clusters with four different masses extracted from the bulk. 
These pdfs  are not flat, on the contrary, they are well described by the function 
\begin{equation}
P(\varphi) = b/(2\pi\sigma^2)^{1/2} \, e^{-(\varphi-\overline\varphi)^2/\sigma^2}
    + (1-b)/(2\pi)
    \; , 
    \end{equation}
with $b, \sigma, \overline\varphi$ fitting parameters. The red  dashed red
line shows one such fit for the data for the largest cluster mass. The extra weight 
close to $\varphi =0$ is more pronounced for the smaller clusters. The particles contributing to 
the peak are the ones for which the active forces point along the direction of motion,
and hence  `push' the cluster. As  the configuration in the inset in Fig.~\ref{fig:quattro}(d) exemplifies,
in this case these particles are situated at the left  boundary and their forces have a preference to point  right, the direction of motion.
Quite typically, the particles with a correlated direction of the active force are situated at the boundaries of the clusters.

Finally, the total torques exerted on the clusters are negligible. The distribution of the ratio between the torque 
acting on an extracted  cluster and its radius of gyration is plotted in Fig.~\ref{fig:quattro}(d). We see that for the two 
Pe values show, the distributions fall off to zero quite rapidly. This measurement is consistent with the fact that 
we do not see the clusters rotate significantly.

\subsection{C6. Distributions of mass variations}
\label{subsec:dist-mass-var}

We now analyze in more detail
the mass variation distributions shown in Fig.~\ref{fig:quattro-new} for the active case at $\rm{Pe}=100$.

\begin{figure}[h!b]
	\centering
	\includegraphics[width=0.75\textwidth]{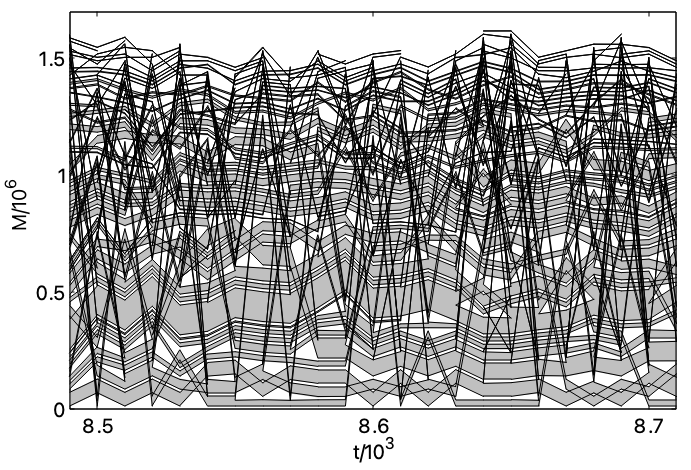}
	\caption{{\bf Tracking of clusters} Graphical representation of the cluster mass evolution in the active system with Pe $=100$, as identified by the tracking algorithm. See the text for a detailed description.
	}
\label{fig:collision_table}
\end{figure}

First, we show in Fig.~\ref{fig:collision_table} an illustrative representation of the result of the tracking of clusters, as identified with the alghoritm described in Sec.~A2. At a given time, each grey band represents, with its width, the mass of one cluster. Moving from one time to the next one, two bands can merge, which corresponds to a coalescence event, or one band can divide into several ones, which corresponds to a breaking event. The cases in which the bands evolve smoothly, isolated from the rest, correspond to clusters which evolve on their own, and they gain or lose mass only through evaporation or condensation of a small number of particles, i.e., Ostwald Ripening.

\begin{figure}[h!]
    \centering
    \includegraphics[scale=0.75]{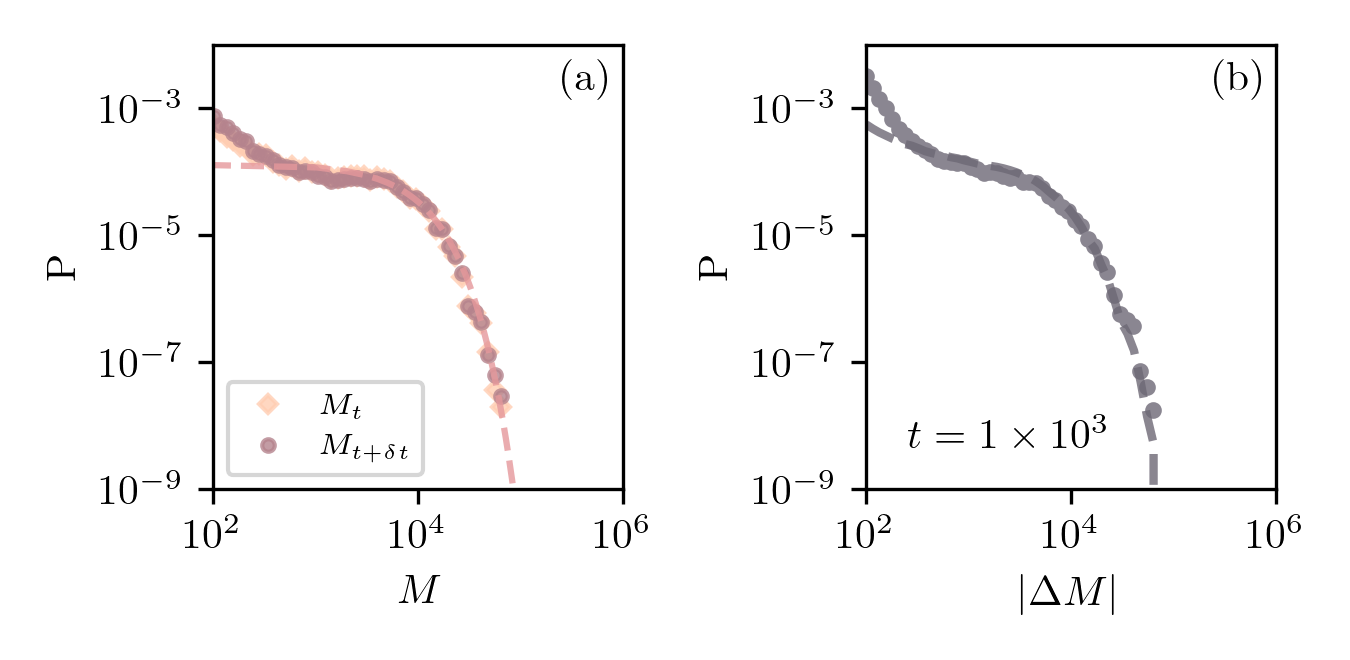}
    \hspace{-0.5cm}
        \includegraphics[scale=0.75]{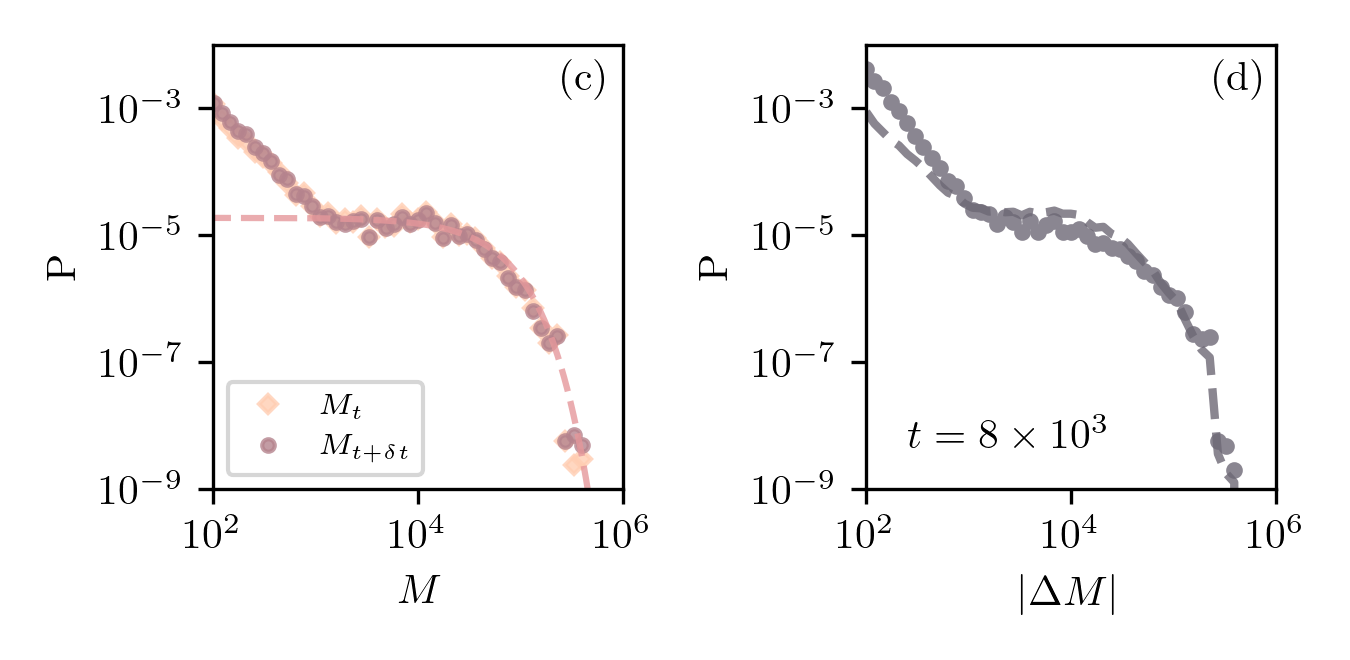}
    \caption{
        {\bf Mass and mass variation distributions at early times.} (a) Distribution of the mass $M_t$ (diamond) and $M_{t+\delta t}$ (circles) for merging/breaking clusters at $t=10^3$, the dashed line shows the fit with an exponential distribution of the 
        large mass tail. (b) Distribution of mass variations $\Delta M = M_{t+\delta t} - M_t$ at $t=10^3$ computed from numerical data (circles), and using Eq.~(\ref{eq:DeltaM}) (dashed line) for the convolution of 
        the numerical probability distributions of $M_t$ and $M_{t+\delta t}$ (see the text for details).  
        (c) and (d) same as in (a) and (b) for $t=8 \cdot 10^3$ 
       The characteristic mass  of  the exponential fits of the tails of the distributions of $M$ at different times is
       $\overline{M}_t \approx 5\cdot10^3 $ and $\overline{M}_t \approx 2\cdot 10^4$ at $t=10^3$ and $t= 8 \cdot 10^3$, respectively. 
       These values are compatible with the mass averages and, within numerical errors, with the  exponential
       tails of the distributions of $|\Delta M|$. 
    }
    \label{fig:dist-m-and-dm-1000}
\end{figure}

We now focus on the variation of mass distributions of clusters undergoing merging and breaking events, that we identify as those with 
 $|\Delta M| \geq 100$.
In Fig.~\ref{fig:dist-m-and-dm-1000}(a) we show the mass distributions $P(M_t)$ and $P(M_{t+\delta t})$ for $t = 10^3$, the starting time 
of the tracking that we choose to be roughly at the beginning of the scaling regime for the corresponding Pe, and $\delta t = 10 \ll t$. We select a 
relatively short $\delta t$ to be able to identify the fate of each cluster correctly.
Since the time delay $\delta t$ is very short compared to $t$, 
the two distributions are practically identical $P(M_t) \sim P(M_{t+\delta t})$. They both 
have a tail that can be well approximated by an exponential decay, as proved in Fig.~\ref{fig:CSDs} and its discussion.

The probability distribution $P_{\Delta M}({\Delta M})$ of the new random variable $\Delta M = M_{t+\delta t}-M_t$, reads
\begin{eqnarray}
    P_{\Delta M} ({\Delta M}) 
    &=& 
    \int dM_t \int dM_{t+\delta t} \; P(M_{t+\delta t}, M_t) \, \delta(\Delta M - M_{t+\delta t} + M_t) 
    \nonumber\\
     &\sim &
     \int dM_t \int dM_{t+\delta t} \; P(M_{t+\delta t}) P(M_t) \, \delta(\Delta M - M_{t+\delta t} + M_t) 
     \nonumber\\
     &=&
     \int dM_t \; P(\Delta M+M_t) \, P(M_t)
    \label{eq:convolution}
\end{eqnarray}
and it just the convolution of the probability density functions of the masses at the two times, if the variables are independent, 
the assumption used in the second line.
If we further use, for the tails, $P(M_{t+\delta t}) \sim e^{-M_{t+\delta t}/\overline M_{t+\delta t}} \sim e^{-M_{t+\delta t}/\overline M_{t}}$
valid for $M_{t+\delta t}>100$,  and similarly for  $P(M_t)$,  the convolution yields another  exponential 
\begin{equation}
    P_{\Delta M} ({\Delta M}) \sim e^{-\Delta M/\overline M_{t}}
    \label{eq:DeltaM}
    \; . 
\end{equation}
In Fig.~\ref{fig:dist-m-and-dm-1000}(a) and (c) we show the probability distribution functions of the 
masses $M_t$ and $M_{t+\delta t}$ at $t = 10^3$ and $t=8 \cdot 10^3$, respectively.  
In panels (b) and (d) 
we compare  the probability distribution of  $\Delta M$ sampled directly with the algorithm 
and the outcome of Eq.~(\ref{eq:convolution}) using the full $P(M_t)$ at the corresponding time $t$. 
The agreement is very good on the full decaying tails, 
and even the small $\Delta M$ regime are rather well captured with this approximation.

\section{D. Single Active Cluster}

In this Section we study the simplest possible model for
a cluster. We ignore the interaction between the cluster and its
environment (gas or other clusters) and we keep the number of particles in the
cluster fixed. Furthermore, for
the sake of simplicity, we consider that the cluster is made of  individual
Active Ornstein-Uhlenbeck Particles (AOUPs) ruled by the
evolution equation
\begin{equation}
    m \dot{\bm{v}}  = - \gamma \bm{v} + \bm{u}_{\rm act} + \bm{\xi}
    \; ,
    \label{eq:single_particle_evolution}
\end{equation}
where $\bm{\xi}$ is a Gaussian white noise with zero mean and $a,b=1,d=2$
component correlations
\begin{equation}
    \langle \xi_a(t) \xi_b(t') \rangle = 2\gamma k_BT \delta_{ab} \delta(t-t')
    \; .
\end{equation}
$\bm{u}_{\rm act}$ is the active force,
modelled as a colored Gaussian noise, described by the equation
\begin{equation}
    \dot{\bm{u}}_{\rm act}  = - \frac{1}{\tau_p}\bm{u}_{\rm act}  +\bm{\eta} \mbox{,}
    \label{eq:act_force_evolution}
\end{equation}
with $\tau_p$ the persistence time and the components $\eta_a$ correlated as
\begin{equation}
    \langle \eta_a(t) \eta_b(t') \rangle = \frac{F_{\rm act}^2}{\tau_p} \, \delta_{ab} \, \delta(t - t')
    \; .
    \label{eq:act_noise_correlation}
\end{equation}
In order to facilitate checks of the expressions we find, we recall that naming $m$, $L$ and $t$ the units of mass, length and time,
the other parameter units are such that
$[m\dot v] = [m] \, L/t^2  = [\gamma] L/t \implies [m/\gamma]=t$. Also, $m L/t^2 = [u_{\rm act}] = [\xi]$
and $[u_{\rm act}]/t = [\eta]$. From the noise correlations $[\xi^2] = [\gamma k_BT]/t$, which implies
$m^2 L^2/t^4 = [\gamma k_BT]/t$. With these units, we find $[k_B T ] = m L^2 / t^2$, which is consistent.
We also have $[\eta] = [F_{\rm act}]/t$. Therefore, $[u_{\rm act}]= [F_{\rm act}]$, as it should.

\begin{figure}[h!]
    \includegraphics[width=\linewidth]{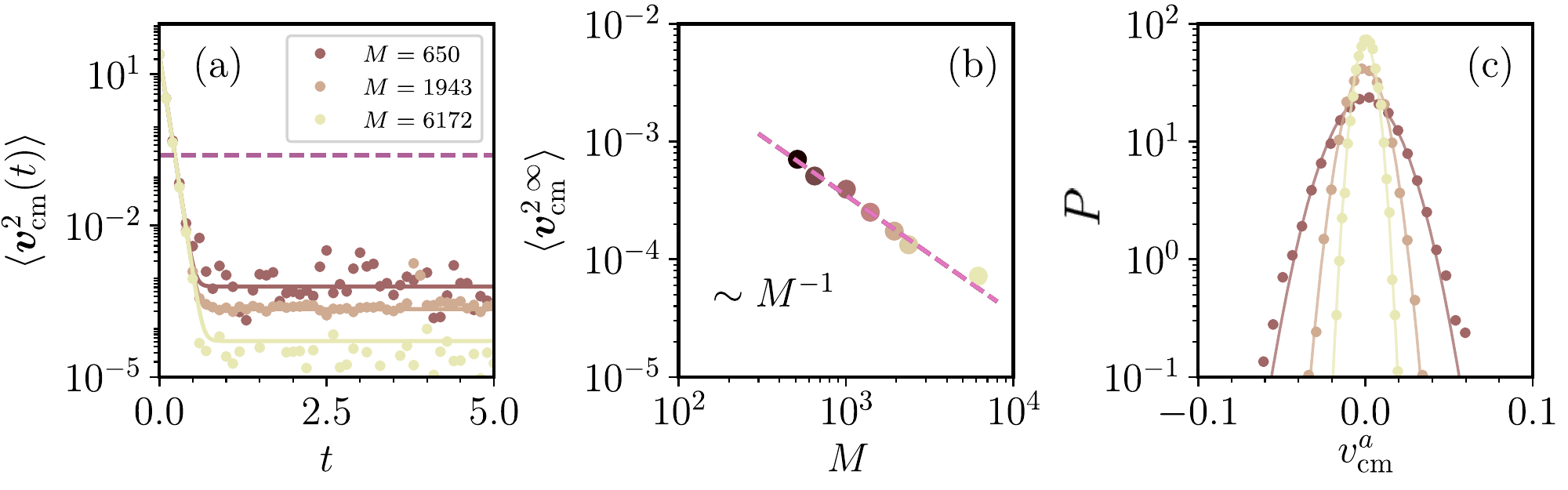}
    \vspace{-0.5cm}
    \caption{ {\bf Comparison between the simple model and the simulations of ABP clusters extracted from the bulk.}
        (a) Squared center of mass velocity of the cluster,
        with an initial velocity term $v_{\rm cm}(0) = 5$.
        Both the first ballistic regime with the single particle characteristic time $m/\gamma$ and
        the plateau proportional to $F_{\rm act}^2/M^2$ at large times are present. The horizontal dashed line
        represents the typical velocity of the single active particles for the same parameters $(F_{\rm act}/\gamma)^2$.
        (b) The asymptotic value $\lim_{t\to\infty} \langle \bm{v}_{\rm cm}^2(t)\rangle$ as a function of the
        cluster mass, together with the law $M^{-1}$.
        (c) Distribution of the center of mass velocities $v_{\rm cm}^a$ along the two Cartesian directions at a fixed
        tracking time for different clusters' masses. The solid lines are fits with Gaussian distributions. Their width
        is related to the clusters kinetic temperature.
    }
    \label{fig:cluster_v2}
\end{figure}

Let's consider a cluster with mass $M$ formed by $M/m$ such AOUPs.
Summing up Eq.~(\ref{eq:single_particle_evolution}) over the particles $i=1, \dots, M/m$
in the cluster $C$, and introducing the quantities
\begin{equation}
    \bm{r}_{\text{cm}} = \frac{1}{M} \sum_{i\in C}m \bm{r}_i
    \; ,
    \qquad\qquad
    \bm{v}_{\text{cm}} = \frac{1}{M} \sum_{i\in C} m \bm{v}_i
    \; ,
\end{equation}
the equation of motion of the center of mass becomes
\begin{equation}
    \dot{\bm{v}}_{\tx{cm}} = - \frac{\gamma}{m} \,  \bm{v}_{\tx{cm}} + \frac{1}{M} \, \bm{U}  + \frac{1}{M} \, \bm{\Xi} \; \mbox{,}
\end{equation}
where $\bm{U} = \sum_{i\in C} {{\bm{u}}_{\rm act}}_i$ and $\bm{\Xi} = \sum_{i\in C} {\bm{\xi}}_{i}$.
The solution is
\begin{equation}
    \bm{v}_{\tx{cm}}(t) = \bm{v}_{\tx{cm}}(0) \; e^{-\gamma t/m}
    + \frac{1}{M} \int_0^t dt_1 \; e^{-(t-t_1)\gamma/m} \; \bm{U}(t_1)
    + \frac{1}{M} \int_0^t dt_1 \; e^{-(t-t_1)\gamma/m } \; \Xi(t_1)
    \; .
\end{equation}
We already see from this expression that the relaxation time of the center of mass velocity, the inertial time,
is the same as the one of a single particle, $m/\gamma$. The short time decay of $\langle \bm{v}^2(t)\rangle$ in
Fig.~\ref{fig:cluster_v2} confirms that this is also the inertial time in the simulation of the ABP clusters extracted
from the bulk.

We want to compare properties of this model to their equivalent in the simulations of the
ABP clusters. In the latter,  we measure the total active force correlation
\begin{equation}
    \bra{\bm{F}(t) \cdot \bm{F}(t')} = 2 F^2 \, e^{-|t-t'|/{\cal T}_p}
    \;, 
    \label{def:correlation}
\end{equation}
see Fig.~\ref{fig:quattro}(a).
The total active force correlation in the ABP clusters decays exponentially with a correlation time ${\cal T}_p$,
which turns out to be, numerically, of the order of the individual $\tau_p$ itself, see Fig.~4(a) in the main text.

The force-force correlation can be expressed in terms of single force correlations as
\begin{equation}
    \label{eq:Fsquared1}
    F^2 \equiv \langle {\mathbf F}_{\rm act}^2(0)\rangle =
    F^2_{\rm act} \, \left[ \sum_{i=1}^{M/m}  \langle {\bf n}^2_i(0) \rangle + \sum_{i\neq j}^{M/m}
        \langle {\bf n}_i(0) \cdot {\bf n}_j(0) \rangle \right]
    \; .
\end{equation}
If one assumes that each particle in the cluster is independent of
the others, and that the ${\bf n}_i$ assume random directions,
\begin{equation}
    \label{eq:Fsquared2}
    F^2  \sim \frac{M}{m} \,  F_{\rm act}^2 \, ( 1 + \zeta)
    \; .
\end{equation}
The $M/m$ dependence represents the data quite accurately, Fig.~\ref{fig:quattro}(b),
with $\zeta$ a number of order one which, actually, takes a quite small value $\zeta \sim 0.05$.
However,  we also know that
this would be an over simplification, given that we observe special alignment of the forces on the boundaries of the
clusters, from time to time, see the inset in Fig.~\ref{fig:quattro}(d). This form gives an idea of the parameter
dependence of $F^2$ which we will adopt in this Section.

Going back to the simple model, the individual noises are truly independent variables and the correlation of the total noise term reads
\begin{equation}
    \bra{\bm{\Xi}(t) \cdot \bm{\Xi}(t')} = \sum_{i \in C} \sum_{j \in C} \sum_{a=1}^d \bra{{\xi}^i_a(t) {\xi}^j_a (t')} = \sum_{ij} \delta_{ij} \;  2 d k_B T \gamma  \,  \delta(t-t') =  \frac{M}{m} \, 4k_B T \gamma \, \delta(t-t')
    \; .
\end{equation}

Using the results for the active force and noise correlations, the two-time velocity correlation becomes
\begin{eqnarray}
    \bra{\bm{v}_{\text{cm}}(t) \cdot \bm{v}_{\text{cm}}(t')}
    &=&
    \bm{v}_{\text{cm}}^2(0) \; e^{-\gamma/m(t+t')}
    \nonumber\\
    &&
    +
    \frac{2F^2}{M^2}
    \
    \frac{1}{(\gamma/m)^2 - 1/{\cal T}_p^2}
    \
    \left[
        \frac{m}{\gamma {\cal T}_p }
        \left(
        e^{-\gamma/m (t+t')}
        -  e^{-\gamma/m |t-t'|}
        \right)
        \right.
        \nonumber\\
        &&
        \left.
        \qquad\qquad \qquad
        +
        \left(
        e^{-\gamma/m (t+t')}
        + e^{-|t-t'|/{\cal T}_p}
        - e^{-t/{\cal T}_p -t'\gamma/m}
        - e^{-t'/{\cal T}_p -t\gamma/m}
        \right)
        \right]
    \nonumber \\
    &&
    + \frac{2k_B T}{M}\left(e^{-\gamma/m(t-t')} - e^{-\gamma/m(t+t')}\right)
    \; .
    \label{eq:v1v2_corr_cluster}
\end{eqnarray}
At equal times
\begin{eqnarray}
    \bra{\bm{v}^2_{\text{cm}}(t)}
    &=&
    \bm{v}_{\text{cm}}^2(0) \; e^{-2\gamma/m \, t}
    \nonumber\\
    &&
    +
    \frac{2F^2}{M^2}
    \  \frac{1}{(\gamma/m)^2 - 1/{\cal T}_p^2}
    \left[
    -
    \frac{m}{\gamma {\cal T}_p }
    \left( 1-  e^{-2 \gamma/m \, t} \right)
    +
    \left(1 + e^{-2 \gamma/m \, t}
    -2e^{-t (1/{\cal T}_p + \gamma/m)}
    \right)
    \right]
    \nonumber
    \\
    &&
    + \frac{2k_B T}{M}\left(1 - e^{-2 \gamma/m \, t}\right)
    \; .
    \label{eq:v1v2_corr_cluster}
\end{eqnarray}
Consistently, one finds  $\bra{\bm{v}^2_{\text{cm}}(0)}
    =
    \bm{v}_{\text{cm}}^2(0)$.
In the long time limit, $t\gg m/\gamma$,
\begin{eqnarray}
    \bra{\bm{v}^2_{\text{cm}}(t)}
    \xrightarrow[t\gg m/\gamma]{}
    \;
    \frac{2F^2}{\gamma^2}
    \,
    \frac{m^2}{M^2}
    \,
    \frac{\gamma {\cal T}_p}{\gamma {\cal T}_p+m}
    + \frac{2k_B T}{M}
    \; .
    \label{eq:modelv2}
\end{eqnarray}
If we now use the parameter dependencies of $F^2$ and ${\mathcal T}_p$ estimated numerically from the
simulations of the ABP model,
\begin{eqnarray}
    \bra{\bm{v}^2_{\text{cm}}(t)}
    \xrightarrow[t\gg m/\gamma]{}
    \;  2 \, ( 1 + \zeta)
    \frac{F_{\rm act}^2 }{\gamma^2}
    \,
    \frac{m}{M}
    \,
    \frac{ \tau_p}{\tau_p+m/\gamma}
    + \frac{2k_B T}{M}
    \; .
\end{eqnarray}
In practice $\tau_p \gg m/\gamma$ and the last factor in the first term is very close to one. Moreover,
$\zeta$ turns out to be much smaller than one. Thus,
\begin{eqnarray}
    \bra{\bm{v}^{2\;\infty}_{\text{cm}}} \equiv \lim_{t\gg m/\gamma} \bra{\bm{v}^2_{\text{cm}}(t)}
    \sim
    \frac{2F_{\rm act}^2 }{\gamma^2}
    \,
    \frac{m}{M}
    + \frac{2k_B T}{M}
    \; .
\end{eqnarray}

From this analysis we conclude that
the equal time center of mass velocity correlation decays over a short timescale related to
the inertia of the single active particles and, at longer times, it reaches a plateau which is inversely proportional
to the cluster mass. This  behavior is confronted to simulations of single clusters extracted from the bulk
in Fig.~\ref{fig:cluster_v2}. In panel (a) we plot the full $\langle \bm{ v}_{\rm cm}^2(t) \rangle$ for three cluster masses,
all of them in the range in which the measured diffusion coefficient decays as $M^{-1/2}$.
The solid lines represent the full analytic expression in Eq.~(\ref{eq:modelv2}).
In panel (b) we plot the asymptotic value $\lim_{t\to\infty} \langle \bm{v}^2(t)\rangle$ as a function of the
cluster mass including more mass values, together with the law $M^{-1}$.
The predictions of the model for this instantaneous measurement are clearly confirmed.
In panel (c) we show the probability density function of the $x$ and $y$ components of the velocity of the center of
mass of the clusters, which we gathered together and simply denoted as $v^a_{\text{cm}}$,
for different values of the clusters' masses. The distributions are well-fitted by
zero-mean Gaussians with a variance that decreases with the mass.

Within the same model,
the cluster diffusion coefficient at long times $t'\gg m/\gamma$, and for $\tau_p \gg m/\gamma$, is
\begin{equation}
    D = \frac{m^2}{M^2} \frac{F^2}{\gamma^2} {\cal T}_p  + \frac{k_B T}{\gamma} \frac{m}{M}
    =
    \left[ (1+\zeta) \,  \frac{F_{\rm act}^2}{\gamma^2} \tau_p  + \frac{k_B T}{\gamma} \right] \frac{m}{M}
    =
    \left[ (1+\zeta) \,  \frac{F_{\rm act}^2}{\gamma^2} \frac{m}{M} \; \tau_p  + \frac{k_B T}{M} \frac{m}{\gamma} \right]
    \; .
\end{equation}
As a consistency check we verify that the units are right. The ones of the last term are clearly correct.
The ones of the first term $[D] = [F_{\rm act}^2]/[\gamma^2] \; [\tau_p] = [u_{\rm act}^2]/[\gamma^2] \; t
    = (mL/t^2)^2/(m/t)^2 \; t = L^2/t$ are also correct. In the limit $m=M$, $\zeta=0$, one recovers the
familiar result for AOUPs and, in the $F_{\rm act} =0$ case.
The more correlated the active forces (higher $\zeta$) the larger the diffusion coefficient.

We note that this simple model predicts a $M^{-1}$ dependence of the diffusion coefficient. This is not what we observe for the
clusters in the bulk, nor for the extracted clusters of not too large mass. The $M^{-1}$ dependence is, asymptotically,
what is measured for the sufficiently large extracted ones, see Figs.~2 and 3 in the main text.
Therefore, the simple model does not capture the parameter dependence of
quantities related to delayed measurements like, for example, the diffusion coefficient. We ascribe this mismatch to two 
reasons. On the one hand, the masses of the active clusters are not fixed and suffer 
from rather large fluctuations during their evolution. This is due to
exchanges with the surrounding gas and clusters and also break-up events. The latter are
exhibited, for example, in Fig. 3 of the main text and movie~M4. On the other hand, the gas made of 
active particles themselves is not just the same as an equilibrium one at an effective temperature. This 
was recently studied in~\cite{Solon22} where the motion of a passive tracer immersed in an active 
dilute bath was analyzed in detail and similar crossover from $D\sim R^{-1}$ to $D\sim R^{-2}$ with 
$R$ the radius of the tracer was found both numerically and with analytic argument.
 The crossover length is controlled by the persistence length of the active particles in the gas and it then 
 approaches zero in the passive limit.

\section{E. The passive attractive case}

In order to better understand the similarities and differences between
active and  passive phase separation we simulated a system
of passive particles ($\rm Pe = 0$) interacting via the same Mie potential $
    U(r)=4\varepsilon [({\sigma}/{r})^{64}-({\sigma}/{r})^{32}]+\varepsilon
$ though with no truncation
at $r=\sigma$ (used in the body of the paper to make it purely repulsive). With its tail, at long distances the
particles feel an attractive interaction.
We chose the density ($\phi = 0.50$) and temperature ($T = 0.30$),
such that the passive attractive system has a liquid-gas co-existence
similar to the  MIPS phenomenology of the ABPs.

\begin{figure}[h!]
    \includegraphics{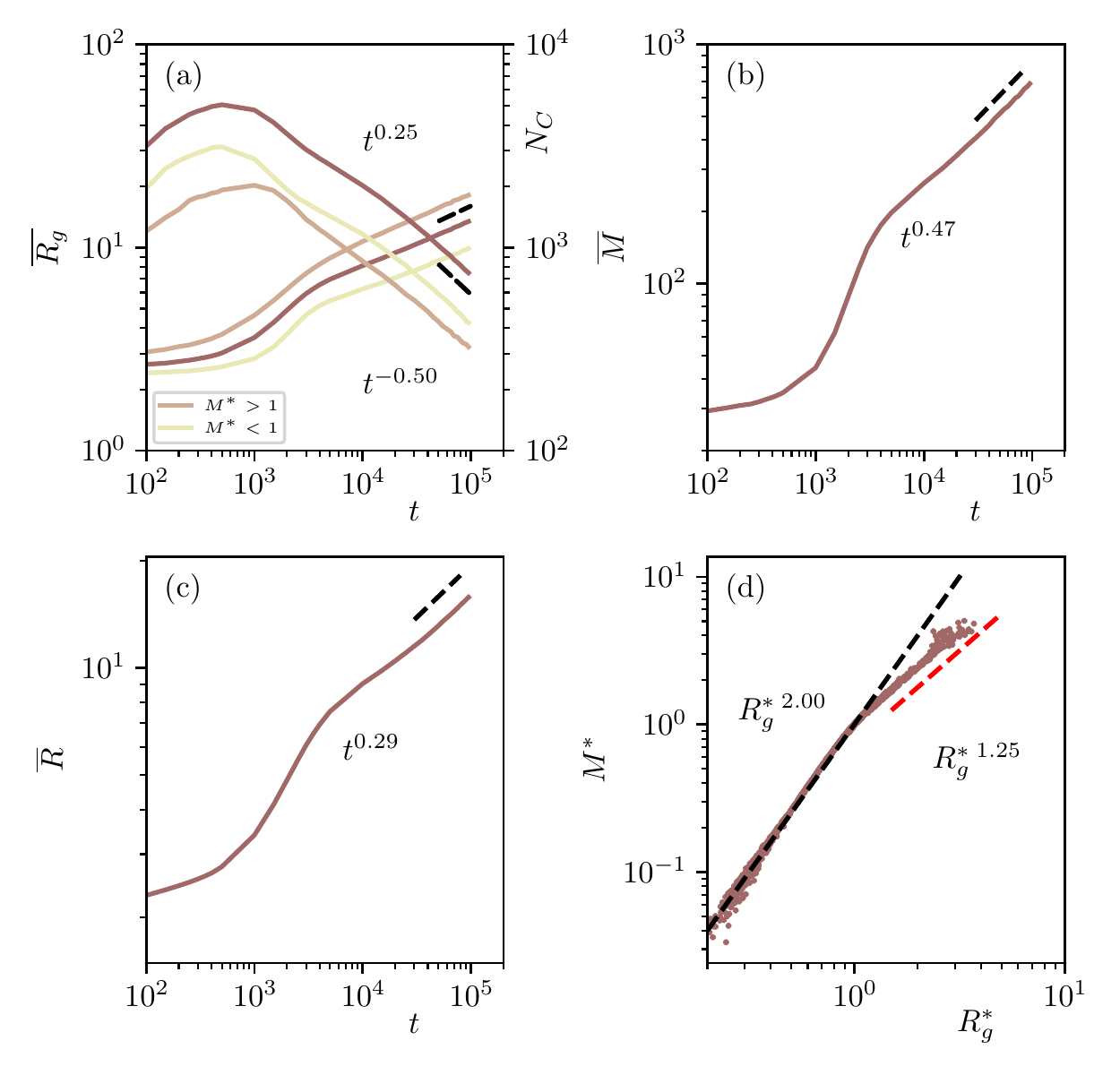}
    \vspace{-0.5cm}
    \caption{{\bf Dynamic behaviour of the attractive passive model.} Time dependence of the averaged cluster radius of gyration in the left scale and
        the number of clusters in the right scale (a),
        the averaged clusters mass (b), the growing length estimated from the scaling properties of the structure factor (c), and
        the scatter plot of $M^*=M/\overline M$ against $R_g^*=R_g/\overline R_g$ (d).
        See the text for a description of the plots. 
    }
    \label{fig:attractive-summary}
\end{figure}

Figure~\ref{fig:attractive-summary} summarizes the main features of the phase separation   in the
attractive case.
Panels~(a) and (b) display the cluster averaged radius of gyration $\overline R_g$ and mass $\overline M$ as functions
of time, together with two dashed lines that indicate the fitted powers in the last reachable regime.
For the sake of comparison, in (c) we plot the growing length as extracted from the scaling properties
of the structure factor. In both cases we see that the system enters the scaling regime at $t_s\sim 5\times 10^3$, considerably
later than in the active model. From the cluster analysis we read
$\overline R_g \sim t^{1/z}$  with $z\sim 4$, which is still far from the expected $z=3$. A similar $z\sim 4$ exponent
was measured in \cite{cerda2004} in a numerical simulation of depletion driven colloidal phase separation. The
study of the structure factor yields an exponent $z$ closer to $z=3$ but still not quite it.
We note that the system could still be exploring a transient regime, before the asymptotic
scaling one. Nevertheless, the times needed to reach the latter are out of our computational
reach for the relatively high densities that we use here in order to compare to the active
case. We note that even in the ferromagnetic Ising model with conserved Kawasaki (local spin exchange) dynamics,
a problem with coarsening dynamics surely characterized by a growing length $R\sim t^{1/3}$,
this asymptotic regime is very difficult to reach~\cite{tartaglia2018coarsening}. The molecular dynamic
simulation of particle systems present even further complications~\cite{watanabe2014ostwald}.

\begin{figure}[h!]
    \includegraphics[scale=0.9]{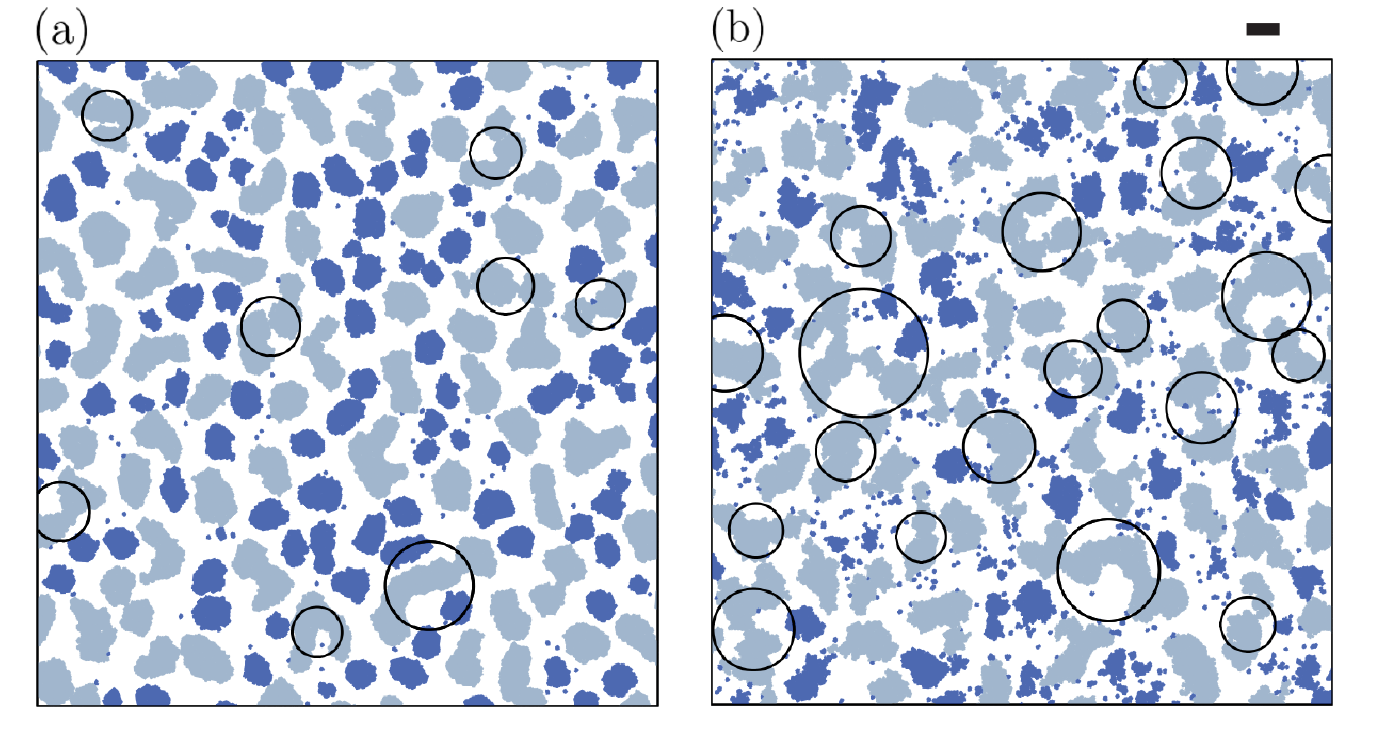}
    \caption{ {\bf Comparison between the configurations of the passive attractive model and the active one.}
        Snapshots  taken from simulations of (a) the passive attractive system described in this section ($\phi=0.5$ and $T=0.3$) and
        (b) the ABPs system described in the main text, with  $\phi=0.5$,  $T=0.05$ and $\rm Pe = 100$. Both configurations have the same
        averaged cluster mass $\overline M$. The average size of the clusters is shown with a line on top of the right panel.
        Fractal clusters are colored in gray ($d_f \approx 1.25$) and regular ones in blue ($d_f=2$). Circles with center 
        at the c.o.m. and radius equal to the gyration radius
        of some of the biggest clusters are also shown, to underscore how fractal clusters are far from having a compact shape.
        The movies M2 and M3 display the evolution of these systems.
    }
    \label{fig:configuration_attractive}
\end{figure}

The scatter plot in (d), $M^* = M/\overline M$ against  $R_g^* = R_g/\overline R_g$
in log-log scale,  is well described by a bi-fractal form, with the exponents $d_f=2$ and $d_f \sim 1.25$.
We note, however, that the extent of points over which we can apply a fit to the second fractal regime
is shorter than in the active case and the spread of data is a bit larger. Though this exponent is
smaller than the 1.45 found for ABP clusters, we cannot assure that it will be the truly
asymptotic one.
Therefore, as in the ABP case, we see small regular clusters with fractal dimension equal to the dimension
of space and a crossover, at the averaged scale $(\overline R_g, \overline M)$, to fractal clusters with a smaller fractal dimension.
This could be confirmed directly by a visual inspection of the configuration,
with an example reported in Fig.~\ref{fig:configuration_attractive}(a).

A  fractal exponent $d_f=1.45$ was measured in~\cite{Paul21} in the cluster analysis of a passive Lennard-Jones system, and in one of the several active models
treated in \cite{alarcon2017morphology}, in its steady state. In these references, no separation of scales was exhibited, and no bi-fractal
structure was shown to exist.

Clusters formed in the passive system have a similar but much more regular shape than the ones formed by MIPS,
which are shown in Fig.~\ref{fig:configuration_attractive}(b).
They are much slower, notably the time needed to enter the scaling regime is around 50 times longer for these
parameters, and this is confirmed by the analysis of the diffusion coefficient
in the main text (cfr. Fig.~\ref{fig:tres}). In particular, their movement is mainly due to fluctuations on
the surface of the cluster, from individual particles leaving/joining the clusters, rather than the active force pushing/breaking the cluster.

\begin{figure}[h!]
    \includegraphics[scale=0.75]{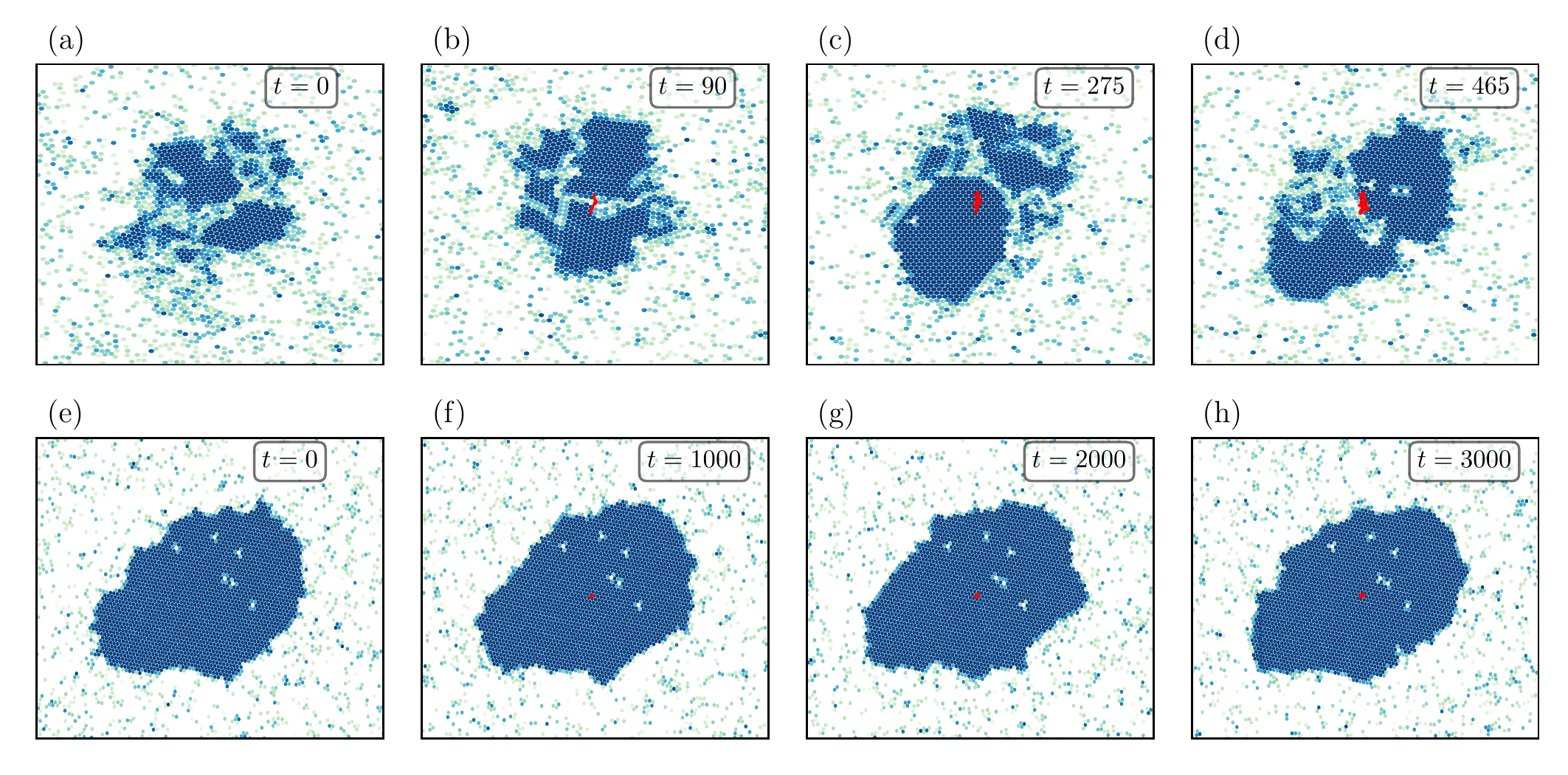}
    \caption{
        {\bf Comparison between the motion of active and passive isolated clusters.}
        Snapshots of (a)-(d) display an isolated cluster of ABPs with $\rm Pe = 100$,
        and (e)-(h) an isolated cluster of passive ($\rm Pe = 0$) attractive particles at the different times indicated in the figures.
        The trajectory of the center of mass of the clusters is represented by a red line. Note the different time scale at which 
        the different snapshots were taken in both cases, clearly showing that the active cluster moves at much smaller time 
        scales than the passive one (which in the time scale shown barely moves at all).
        A comparison between the histories of the two
        clusters is also exposed by the movie M4.}
    \label{fig:comparison_attr_active}
\end{figure}

A direct comparison of the motion of the  clusters in the ABP systems and the passive one is reported in Fig.~\ref{fig:comparison_attr_active} where we
confront the evolution of clusters taken out of the full system, and set in contact with a gas that mimics their conditions in the bulk.
In particular,
in Fig.~\ref{fig:comparison_attr_active}(a)-(d), the time evolution of a single ABPs cluster is shown for $\rm Pe = 100$.
One can see that the evolution is characterized by dramatic fluctuations in shape and mass. In general, due to the non-attractive
nature of the clustering, breaking events are frequent. In panels (e)-(h), a cluster of attractive particles ($\rm Pe = 0$) of comparable
size and mass,
is shown. In this case, the structure of the cluster remains the same along the time evolution and, in general, the cluster is stable,
there are no grain boundaries inside and dramatic breaking is extremely rare. The movies M3 and M4 illustrate these
issues very clearly.

\begin{figure}[h!]
    \includegraphics[scale=1.0]{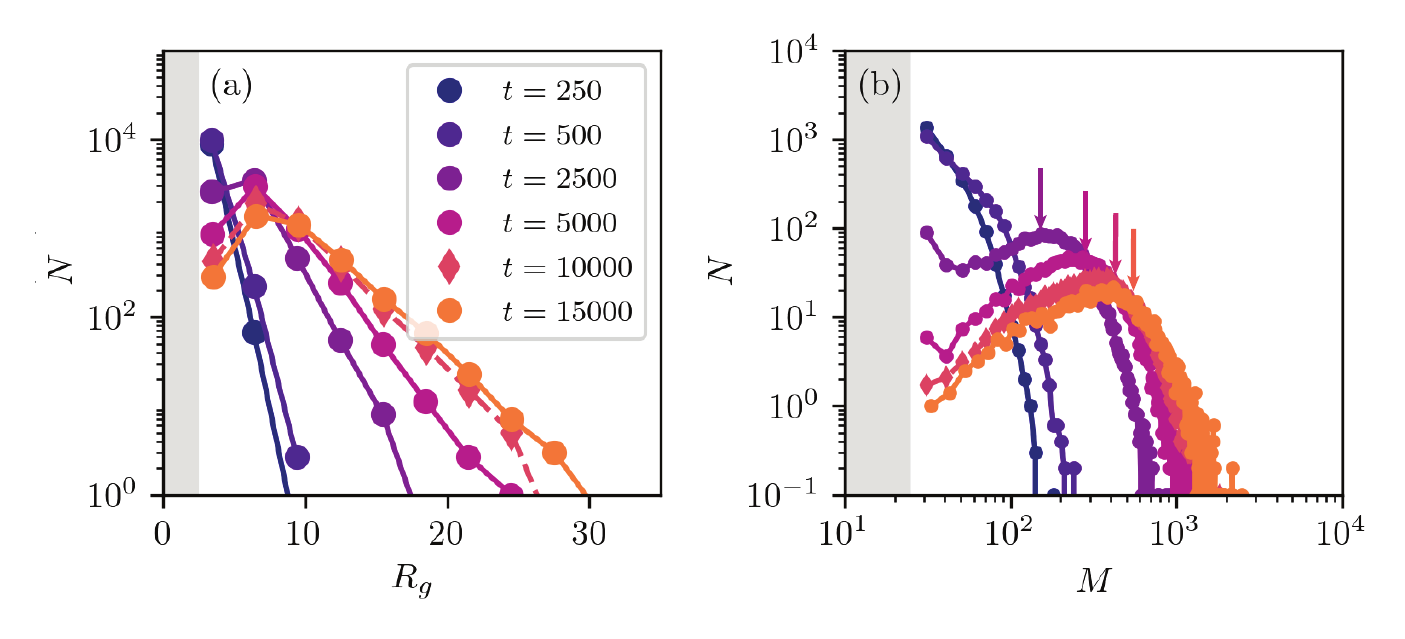}
    \vspace{-0.5cm}
    \caption{{\bf The distribution of cluster sizes in the passive attractive model}, at different times given in the key.
        The gray regions are excluded by the cut-off $M>25$ imposed on the size of the clusters. The
        data points plotted in violet scale are for very short times, well before the scaling regime, $t \ll t_s$. The
        data points  in red-dark orange are for times in the intermediate regime of rapid increase of $\overline R_g$.
        Finally, the two last sets of data (orange) are in the scaling regime, $t > t_s$. The vertical arrows in (b) indicate the averaged values.
    }
    \label{fig:dist-Rg-attractive}
\end{figure}

Figure~\ref{fig:dist-Rg-attractive} displays the probability distribution of cluster sizes in the passive attractive model.
These curves are very different from the ones we measured in the active system, shown above (see Fig. \ref{fig:CSDs}). In particular, in the scaling
regime, after a very fast exponential decay of the weight of small clusters, the distribution has a clear minimum followed by
a maximum at two well-defined sizes. For even  larger clusters, the decay is close to exponential again.
The non-monotonic form of $N(R_g)$ excludes a double
exponential as a possible description of it.

The corresponding
histograms of the cluster masses are displayed in the right panel of the same figure. One can verify that they also show
differences with  the active cluster mass distributions plotted in Fig.~\ref{fig:CSDs}. The arrows indicate the
average values $\overline M$. While here one clearly distinguishes the minimum-maximum structure, in the active model
this is replaced by a plateau in the same double log representation.

\begin{figure}[h!]
    \includegraphics[scale=1.0]{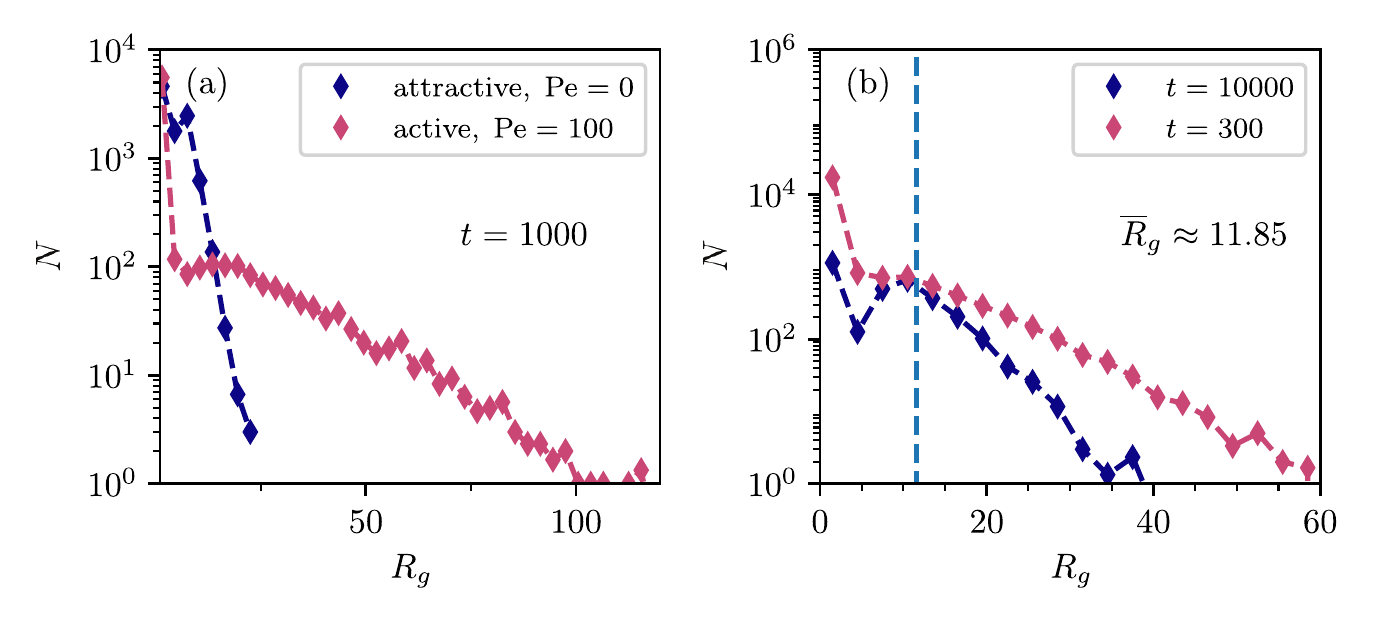}
    \vspace{-0.5cm}
    \caption{{\bf The distribution of cluster radii of gyration in the active and passive models.} In (a) data are
        taken at the same time $t$, while in (b) they were gathered at different times such that $\overline R_g$ is the
        same.  The color of the data points is given by the active or passive character of the  system and is the
        same in the two panels.  The vertical dashed line in (b) indicates the averaged ${\overline R}_g$ which is the same in
        the two systems.
    }
    \label{fig:SM-comparison}
\end{figure}

An interesting question now comes to mind: how do these distributions compare to the ones of the same
quantities in the active model? In Fig.~\ref{fig:SM-comparison}(a) we plot the number  of radii of
gyration in the two systems at the same time while in Fig.~\ref{fig:SM-comparison}(b) we plot the same
numbers at times such that the averaged $\overline R_g$ are the same, and is indicated by the vertical dashed line.
In both figures  the tail of the
active case is longer and higher than in the passive case, proving that the active system has more fractal
clusters.

\begin{figure}[h!]
    \includegraphics[scale=0.9]{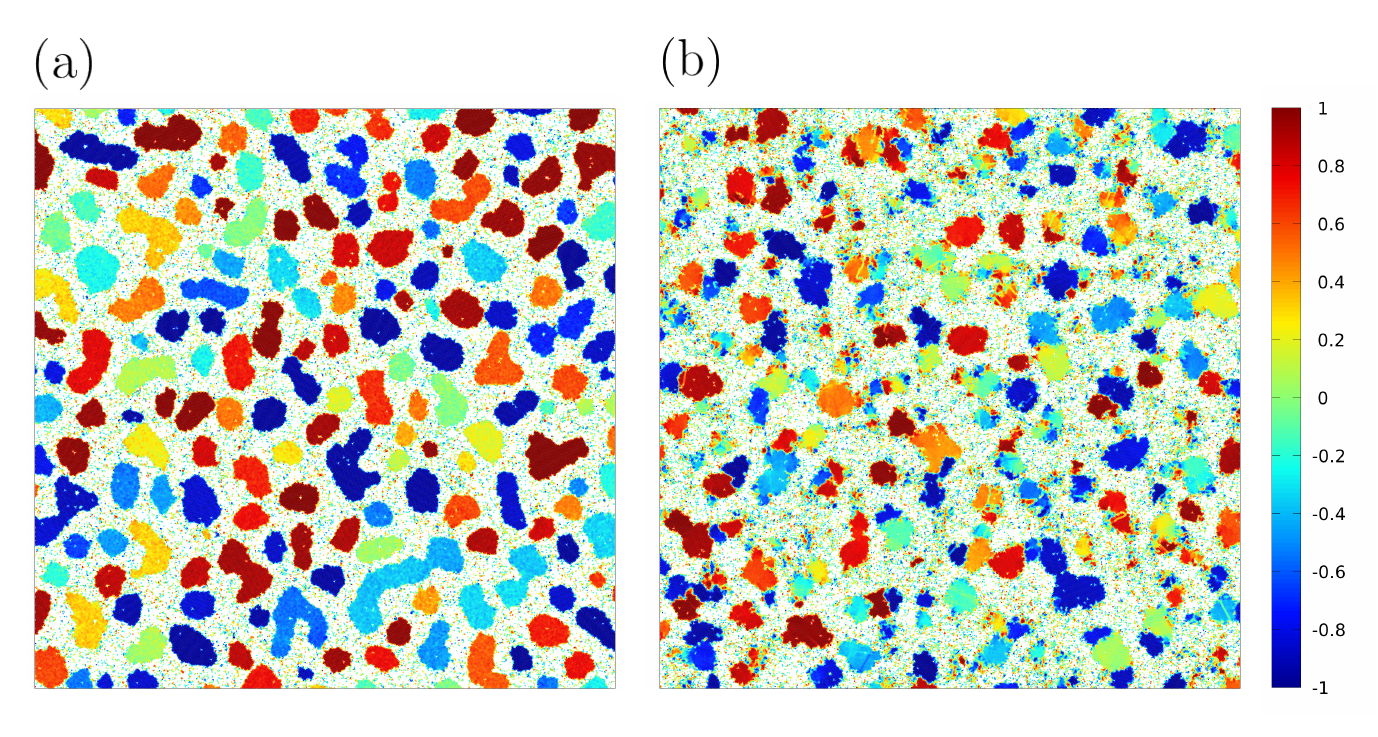}
    \caption{{\bf Local hexatic order.} Same configurations as in Fig.~\ref{fig:configuration_attractive}, but with particles
        colored according to the value of the projection of the local hexatic order parameter
        over the global
        average, see \cite{PRLino} for details. In the attractive passive case, clusters
        are formed by single hexatic domains, while in the active case, large clusters exhibit a micro-separation
        in smaller hexatic patches~\cite{Caporusso20}.
    }
    \label{fig:hexatic_attractive}
\end{figure}

Finally, while in the active case we see that large clusters are formed by smaller patches with different
orientational order, in the passive case the dense clusters have a uniform orientational
order. These facts can be observed in Fig.~\ref{fig:hexatic_attractive} where we plot the same
configurations of Fig.~\ref{fig:configuration_attractive} using different colors for different orientational
orders, as quantified by the local hexatic order parameter,
\begin{equation}
    \psi_i = \frac{1}{nn_i} \sum_{j \in \partial_i} e^{6 {\rm i} \theta_{ij}}
\end{equation}
where the sum runs over the $nn_i$ first neighbors of the $i$th particle defined via a Voronoi tesselation,
and $\theta_{ij}$ is
the angle formed by the segment that connects the center of the $i$th disk with the one of its $j$th
nearest neighbor and the horizontal axis.

Video M4 illustrates what is going on in phase separation of the
passive attractive system. On the one hand,
some small clusters evaporate and single particles from the gas get attracted to larger ones, according to the Ostwald ripening
mechanism. On the other hand,
there is some cluster aggregation. However, differently to what happens in the active case, here the clusters
collide, detach and rotate a little bit, until they join in such a way that the whole new cluster gets a global homogeneous
hexatic order. Some examples of these processes are clearly seen in the video. The larger clusters thus formed are
elongated and therefore have a rather small fractal dimension. In the course of time, if not hit by another
cluster, they will progressive get more round shaped.

\begin{figure}[h!]
    \includegraphics{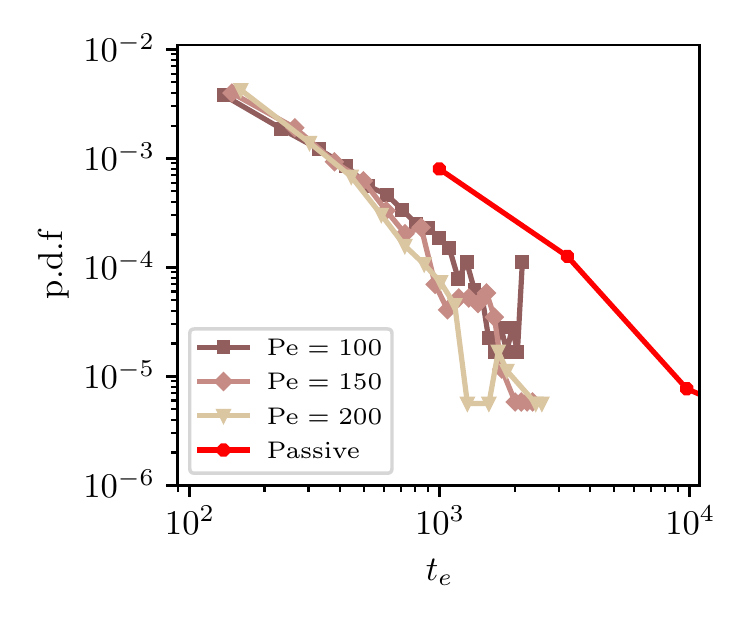}
    \vspace{-0.5cm}
    \caption{{\bf Distribution of tracking end-times} $t_e$ for the active and passive models.
    }
    \label{fig:end-times}
\end{figure}

We finally stress that in the passive system the time tracking of clusters, as described in Sec.~\ref{sec:tracking}, lasts much longer than in the
active one. This is due to the fact that the passive clusters are much more stable
than the active ones and there are less collisions in the former than in the latter case.
This statement can be made quantitative by studying the distribution of end-times $t_e$.
The end-times are the times at which tracking stops, according to the
criterium explained in Sec.~A2. It is quite clear from Fig.~\ref{fig:end-times}  that in
the passive model the distribution has a much slower decay.

\vspace{0.25cm}

We summarize here the outcome of the comparison between active and passive clusters.
\begin{itemize}
    \item
          Clusters of passive attractive particles move more slowly than the active ones: the diffusion constant decays faster with $M$ and its
          absolute value is around 40 times smaller for the parameters that we focus on (Fig. 3 of the main text).
    \item
          The large clusters of active particles have patches with different hexatic order within (micro phase separation under way)
          while the ones of passive particles are uniform in the computational time scales.
    \item
          The distributions of gyration radii and masses are different in active and passive systems.
    \item
          In the passive attractive case we see mostly evaporation of small clusters and single particle attachment to larger clusters (Ostwald ripening). There is, though, some collision of rounded clusters that gives rise to very elongated clusters with rather low $d_f$.
    \item
          In the active model we see more merging of clusters and less evaporation. In a sense the ordering process is
          closer to a Smoluchowski diffusion-aggregation one.
    \item
          The clusters interfaces are much rougher in the active than in the passive case, where they look pretty smooth.
    \item
          In both systems there is co-existence of regular and fractal clusters. For times such that the averaged cluster sizes are the same,
          the regular ones are more numerous than the fractal (by a factor 3, approximately).
\end{itemize}













\end{document}